\documentclass[sigplan,screen]{acmart}
\startPage{1}

\setcopyright{cc}
\setcctype{by-nc-nd}
\acmDOI{10.1145/3774934.3786454}
\acmYear{2026}
\copyrightyear{2026}
\acmISBN{979-8-4007-2310-0/2026/01}
\acmConference[PPoPP '26]{Proceedings of the 31st ACM SIGPLAN Annual Symposium on Principles and Practice of Parallel Programming}{January 31 -- February 4, 2026}{Sydney, NSW, Australia}
\acmBooktitle{Proceedings of the 31st ACM SIGPLAN Annual Symposium on Principles and Practice of Parallel Programming (PPoPP '26), January 31 -- February 4, 2026, Sydney, NSW, Australia}

\usepackage{algorithm}
\usepackage{algpseudocode}
\algnewcommand{\LineComment}[1]{\State \(\triangleright\) #1}
\usepackage{booktabs} %
\usepackage{tabularx}
\usepackage{array}
\usepackage{url}
\usepackage{multirow}

\usepackage{dblfloatfix}
\usepackage{cuted}

\usepackage{listings}

\usepackage{pgfplots}
\pgfplotsset{compat=1.18}
\usepgfplotslibrary{groupplots}
\usepgfplotslibrary{colorbrewer}
\usepackage[table]{xcolor}
\usepackage{xcolor}
\usepackage{listings}
\definecolor{mGreen}{rgb}{0,0.6,0}
\definecolor{mGray}{rgb}{0.5,0.5,0.5}
\definecolor{mPurple}{rgb}{0.58,0,0.82}
\definecolor{backgroundColour}{rgb}{0.95,0.95,0.92}
\lstdefinestyle{CStyle}{
    backgroundcolor=\color{backgroundColour},   
    commentstyle=\color{mGreen},
    keywordstyle=\color{magenta},
    numberstyle=\tiny\color{mGray},
    stringstyle=\color{mPurple},
    basicstyle=\footnotesize,
    breakatwhitespace=false,         
    breaklines=true,                 
    captionpos=b,                    
    keepspaces=true,                 
    numbers=left,                    
    numbersep=5pt,                  
    showspaces=false,                
    showstringspaces=false,
    showtabs=false,                  
    tabsize=2,
    language=C
}
\usepackage{pgfplotstable}
\usepackage{balance}
\usepackage[utf8]{inputenc}
\usepackage{microtype}
\usepackage{siunitx}
\usepackage{xurl}

\begin{document}

\title[Dynamic Detection of Inefficient Data Mapping Patterns...]{Dynamic Detection of Inefficient Data Mapping Patterns in Heterogeneous OpenMP Applications}

\author{Luke Marzen}
\orcid{0009-0001-6791-7732}
\affiliation{%
  \institution{Iowa State University}
  \city{Ames}
  \state{Iowa}
  \country{USA}
}
\email{ljmarzen@iastate.edu}

\author{Junhyung Shim}
\orcid{0009-0003-3899-5488}
\affiliation{%
  \institution{Iowa State University}
  \city{Ames}
  \state{Iowa}
  \country{USA}
}
\email{jshim@iastate.edu}

\author{Ali Jannesari}
\orcid{0000-0001-8672-5317}
\affiliation{%
  \institution{Iowa State University}
  \city{Ames}
  \state{Iowa}
  \country{USA}
}
\email{jannesar@iastate.edu}

\begin{abstract}

With the growing prevalence of heterogeneous computing, CPUs are increasingly being paired with accelerators to achieve new levels of performance and energy efficiency.
However, data movement between devices remains a significant bottleneck, complicating application development.
Existing performance tools require considerable programmer intervention to diagnose and locate data transfer inefficiencies.
To address this, we propose dynamic analysis techniques to detect and profile inefficient data transfer and allocation patterns in heterogeneous applications.
We implemented these techniques into \texttt{OMPDataPerf}, which provides detailed traces of problematic data mappings, source code attribution, and assessments of optimization potential in heterogeneous OpenMP applications.
\texttt{OMPDataPerf} uses the OpenMP Tools Interface (OMPT) and incurs only a \SI{5}{\%} geometric‑mean runtime overhead.

\end{abstract}




\maketitle

\section{Introduction}

The shift towards heterogeneous computing in HPC, where CPUs are paired with accelerators, often GPUs, has enabled new levels of power efficiency and absolute performance.
However, harnessing this unprecedented computational power requires programming models that can manage the complexities of offloading computations to GPUs while maintaining ease of use and portability.
OpenMP has become the most widely adopted parallel programming API \cite{kadosh2023} and has evolved to support GPU offloading, enabling developers to leverage the computational power of GPUs with minimal changes to their existing codebases.
OpenMP offloading provides a standardized, portable, and user-friendly approach for developing high-performance applications for heterogeneous environments.
However, this paradigm shift brings forth challenges in efficiently managing data movement across diverse computing resources, as each connected device typically has its own distinct memory space. 
Data transfers between CPUs and their attached accelerators have higher latency and lower throughput than accesses to directly attached memory.
To maximize performance, programmers explicitly define data mappings between devices using OpenMP's \texttt{target data} and \texttt{target update} constructs.
This requires careful consideration of data flow and validity across multiple memory spaces, a process that is often laborious and time-consuming in real-world codebases.
Nonetheless, minimizing data movement remains a key optimization goal, as data movement directly impacts execution time and, if left unoptimized, can potentially outweigh any performance benefits gained from offloading computations.

Due to the critical importance of this task for performance and correctness, previous works and even extensions of the OpenMP standard have sought to alleviate this burden.
OMPSan \cite{barua2019} detects bugs related to incorrect data mapping through static analysis.
OMPDart \cite{marzen2024static} introduced static code analysis techniques that can be used to generate efficient data map clauses.
Additionally, various compile-time optimizations have been proposed to reduce redundant data transfers \cite{barua2020, guo2023}.
Archer \cite{archer}, Arbalest \cite{yu2021}, and Arbalest-Vec \cite{Arbalest-Vec} use dynamic analysis techniques to identify data races in OpenMP programs.
In the 5.0 revision of the OpenMP specification \cite{openmpSpec50}, support for {\it Unified Shared Memory} (USM) was introduced, which relies on runtime implementations to manage data transfer.
However, USM is heavily dependent on compiler, driver, and hardware integration. 
Current implementations exhibit performance portability and/or stability issues \cite{elwasif2023, mishra2017}, which continue to deter application developers from adopting this approach despite its potential to reduce development effort.

In this work, we present dynamic analysis methods for detecting and profiling data transfer and memory allocation patterns that can bottleneck performance in heterogeneous programs.
These problematic data mapping patterns, outlined in Section \ref{sec:problematic_patterns}, are currently \emph{not} explicitly reported by existing performance tools, including NVIDIA NSight System \cite{nsys}, AMD ROCProfiler \cite{rocprofiler}, Intel VTune Profiler \cite{intel-vtune}, and Intel Advisor \cite{intel-advisor}; instead these tools leave it to the programmer to infer whether or not optimization potential exists based on the coarse-grained timing and transfer volume information and visualizations that these tools provide.
We introduce {\tt OMPDataPerf}\footnote{Source code available at \url{https://github.com/lmarzen/ompdataperf} and \url{https://zenodo.org/records/18102356}.}, a compiler- and hardware-agnostic dynamic analysis tool designed to identify inefficient data mapping patterns, profile them, and provide actionable feedback with estimations of performance uplift if the identified issues are eliminated.

The main contributions of this work include:
\begin{itemize}
    \item low-overhead dynamic analysis methods for identifying inefficient data transfer and memory allocation patterns in heterogeneous applications.
    \item \texttt{OMPDataPerf}, a portable, low-overhead performance tool for detecting, profiling, and providing source code attribution for inefficient data mapping patterns in OpenMP programs.
    Additionally, \texttt{OMPDataPerf} accurately estimates performance uplift if the identified issues are fixed which can guide optimization efforts.
    \item fixes for performance issues identified in programs including speedups of \SI{110}{\%} in Rodinia's \texttt{bfs}, \SI{7}{\%} in Mantevo's \texttt{minife}, and \SI{14}{\%} in bspline-vgh-omp in \texttt{HecBench}.
\end{itemize}

\section{Background} \label{sec:background}

\subsection{OpenMP Offload Terminology}

Throughout this paper, we use OpenMP-centric terminology.

\textit{Device}:
``An implementation-defined logical execution engine'' \cite{openmpSpec52}.
This could correspond to a CPU, GPU, or other types of processing units.

\textit{Host Device}:
``The device on which the OpenMP program begins execution'' \cite{openmpSpec52}.
This refers to the system's main processor (CPU).

\textit{Target Device}:
``A device with respect to which the current device performs an operation, as specified by a device construct or an OpenMP device memory routine'' \cite{openmpSpec52}.
The target device is typically a different hardware unit, such as a GPU or accelerator, that is intended to execute parallel computations efficiently.

For succinctness, the {\it host device} is often referred to as the {\it host} and {\it target device(s)} as {\it device(s)}.

\subsection{OpenMP Offloading Execution Model}

OpenMP offloading enables programmers to more productively leverage the compute capabilities of modern heterogeneous infrastructures through high-level annotations for parallelizing and offloading computations from CPUs to accelerators.
Since the introduction of offloading in OpenMP 4.0 \cite{openmpSpec40} and subsequent improvements in OpenMP 4.5 \cite{openmpSpec45}, the growing availability of OpenMP offload capable compilers has reduced reliance on vendor-specific compiler infrastructures and specialized kernel languages, simplifying the use of GPUs for general-purpose computing.
Offloading with OpenMP consists of two main steps: (i) mapping and transferring data between the host and the device and (ii) executing annotated computations on the device.
OpenMP uses \texttt{target} directives to indicate that a specific code region should be offloaded to an accelerator.
Within these \texttt{target} regions, one or more teams of threads are executed on a device. \texttt{target} regions have an implicit data environment that can be explicitly managed and manipulated with the use of \texttt{map} clauses for fine-grain control of data transfer.

\subsection{OpenMP Tools Interface}
\label{subsec:ompt}
The {\it OpenMP Tools Interface} (OMPT) \cite{ompt-tr, ompt, ompt-tr2}, was standardized in the OpenMP 5.0 specification \cite{openmpSpec50} with the goal of enabling portable tools for analysis of OpenMP applications.
OMPT is an API integrated into the OpenMP runtime, which allows tools to monitor and trace OpenMP events as they occur during execution.
Leveraging integration into the runtime brings the advantage that the instrumentation is independent of the user's code.
Furthermore, tools built using OMPT can operate across different (compliant) OpenMP implementations, enabling them to be hardware- and compiler-agnostic.
However, some tools may depend on optional features not provided in minimal OMPT implementations.

OMPT provides mechanisms for registering callbacks, tracing activity on target devices, and examining OpenMP state.
Callbacks enable tools to receive notification of and offer visibility into the execution of OpenMP events such as parallel regions, work-sharing constructs, synchronization, task execution, and target operations.
The target tracing interface exists to enable the monitoring activities on targets in cases where target devices do not initialize a full OpenMP runtime system which may render callbacks not possible.

OMPT was extended in OpenMP 5.1 \cite{openmpSpec51}, bringing the External Monitoring Interface (EMI) callbacks in addition to a few minor revisions.
The EMI callbacks provide notifications at both the start and end of target events, whereas the non-EMI versions only notify at the start of the events and are now deprecated as of OpenMP 6.0 \cite{openmpSpec60}.

\section{Motivation} \label{sec:motivation}
There have been research works, such as DrGPUM \cite{DrGPUM}, ValueExpert \cite{ValueExpert}, that aim to detect inefficient data transfers through heavy instrumentation of CUDA binaries. Although these are great profilers, they are vendor-specific and thus limit their use for other vendors' accelerators.

As mentioned in section \ref{subsec:ompt}, OMPT is hardware agnostic, which allows for hardware-independent analysis of data transfers. Despite providing a great utility, there is not a tool for checking inefficient data movements using OMPT interface to our knowledge. The closest state-of-the-art tool that utilizes OMPT is Arbalest-Vec \cite{Arbalest-Vec}. However, this is a data correctness checking tool that checks for \textit{data races}, \textit{use of uninitialized memory (UUM)}, \textit{use of stale data (USD)}, \textit{use after free (UAF)}, and \textit{buffer overflow (BO)}. We later show in section \ref{sec:evaluation_comparative_study} that report of these errors alone may not facilitate programmers to write efficient code. In recognition of these problems with existing tools, we developed \texttt{OMPDataPerf} to equip programmers with a hardware-agnostic profiler that detects inefficient data movements. 

\section{Inefficient Data Mapping Patterns} \label{sec:problematic_patterns}

In this section, we provide an overview of four categories of inefficient data mapping patterns that we aim to detect.
Inefficient mappings may arise when data already resides in memory on the destination device, when memory is reallocated on target devices despite sufficient capacity, or when mapped data goes unused.
These issues commonly manifest in OpenMP offload programs as the result of programmers not explicitly mapping data over multiple kernel executions, failing to recognize data reuse opportunities, and other programming oversights.

Joubert et al. \cite{joubert2015} identified data reuse and minimizing memory traffic as essential performance considerations for large-scale scientific applications.
Mishra et al. \cite{mishra2020} evaluated the performance of 10 OpenMP offload benchmarks and found that incorporating data reuse can substantially reduce program execution time compared to the same benchmarks without data reuse.
Marzen et al. \cite{marzen2024static} improved data reuse in the OpenMP offload version of LULESH 2.0 \cite{lulesh} to achieve a \SI{1.6}{\times} speedup in overall execution time over the existing implementation.
Welton et al. \cite{welton2018} found that duplicate data transfers increased runtime by \SI{40}{\%} in QBox \cite{qbox} and QBall \cite{qball}.
The inefficient mapping patterns we have identified are symptoms of missed data reuse opportunities and generally bad data mapping practices.
These patterns signify the presence of potential performance issues and are \emph{not} indicative of correctness issues.
For each pattern, we provide precise definitions and illustrate them with straightforward examples.

\subsection{Duplicate Data Transfers}

\begin{definition} \label{def:duplicate_data_transfers}
A \textit{duplicate data transfer} occurs when a device (or host) receives data that it had previously received.
\end{definition}

Duplicate data transfers can bottleneck program performance, especially when the transfers are repeated frequently in large amounts of data.
Listing \ref{lst:motivation_1} depicts duplicate data transfer between kernel executions.
In this example, the programmer failed to recognize the opportunity to reuse array \texttt{a} for the second target region.
To fix this, array \texttt{a} could be mapped over both target regions using a \texttt{target data} directive.

\begin{lstlisting}[
  label=lst:motivation_1,
  caption={Duplicate data transfer occurs since \texttt{a} is transferred to the device before entering each target region. Required device memory is also allocated and deallocated for each target region.},
  captionpos=b,
  language=C,
  otherkeywords={define,pragma,\# },
  numbers=left,
  stepnumber=1,
  tabsize=2,
  showspaces=false,
  showstringspaces=false,
  basicstyle=\ttfamily\footnotesize,
  xleftmargin=20pt,
  float=htb,  %
  belowskip=-\parskip,
]
int a[N], sum = 0, prod = 1;
#pragma omp target map(to:a) map(tofrom:sum)
#pragma omp parallel for reduction(+:sum)
for (int i = 0; i < N; ++i) {
    sum += a[i];
}

#pragma omp target map(to:a) map(tofrom:prod)
#pragma omp parallel for reduction(+:prod)
for (int i = 0; i < N; ++i) {
    prod *= a[i];
}
\end{lstlisting}

\subsection{Round-Trip Data Transfers}

\begin{definition} \label{def:round_trip_data_transfer}
A \textit{round-trip data transfer} occurs when a device (or host), $A$, sends data to another device, $B$, and later device $A$ receives the same unmodified data back from device $B$.
\end{definition}

Listing \ref{lst:motivation_2} shows a kernel that is called repeatedly inside a loop.
Since no explicit data mappings are present, the implicit data-mapping rules are applied.
After each iteration, the intermediate result is stored in array \texttt{a} and copied from the device to the host.
At the beginning of the next iteration, array \texttt{a} is copied back to the device without modification, hence completing a round-trip.
This code snippet could be optimized by adding a data mapping directive over the outer loop to eliminate the round-trip transfers between kernel executions.

\begin{lstlisting}[
  label=lst:motivation_2,
  caption={A kernel nested inside a loop without a data mapping directive before the outer for loop results in round-trip data transfer each iteration of the outer loop. This example also exhibits repeated device memory allocations since array \texttt{a} is reallocated every iteration.},
  captionpos=b,
  language=C,
  otherkeywords={define,pragma,\# },
  numbers=left,
  stepnumber=1,
  tabsize=2,
  showspaces=false,
  showstringspaces=false,
  basicstyle=\ttfamily\footnotesize,
  xleftmargin=20pt,
  float=htb,  %
  belowskip=-\parskip,
]
int a[N] = {};
for (int i = 0; i < N; ++i) {
    #pragma omp target parallel for
    for (int j = 0; j < N; ++j) {
        a[j] += j;
    }
}    
\end{lstlisting}

\subsection{Repeated Device Memory Allocations}

\begin{definition} \label{def:repeated_device_memory_allocation}
A \textit{repeated device memory allocation} occurs when memory on a target device is allocated, and subsequently deleted (deallocated), more than once to accommodate the mapping of the same variable.
\end{definition}

Repeated device memory allocation and deletion can introduce unnecessary overhead and should be avoided.
This is typically caused when programmers use mappings whose lifetime does not extend across multiple kernels, which causes allocation and deletion at the start and end of each kernel.
Reallocation is sometimes exhibited as a symptom of duplicate and round-trip data transfers and is present in the simple examples shown in Listings \ref{lst:motivation_1} and \ref{lst:motivation_2}.
This behavior can also happen without the presence of duplicate and round-trip transfers, such as in cases where data is modified between transfers.
Reallocating and deleting device data is occasionally done purposefully in cases when GPU memory capacity would be insufficient to allocate all data over the program's lifetime, but should otherwise be avoided.

\subsection{Unused Data Mappings}

\begin{definition} \label{def:unused_data_mapping}
An \textit{unused data mapping} occurs when data is mapped to a device, either by copying data to the device or by allocating memory on the device, but the device does \emph{not} read the copied data or utilize the allocated region during the lifetime of the mapping.
\end{definition}

Unused data mappings are sometimes introduced into programs that contain dead code, overly cautious preemptive transfers, or conditional logic that sometimes bypasses kernel execution.
This is almost always unintentional and is certainly bad practice.

\section{Detection} \label{sec:detection}

In this section, we present a set of algorithms to identify and categorize inefficient data transfer patterns across an arbitrary number of devices.
These categorizations, along with precise source code locations, can help programmers to quickly diagnose the root causes of data movement inefficiencies.

It is an important design consideration for performance profiling tools that their impact on execution time is minimized to preserve accurate and meaningful performance metrics and reduce the need for error-prone overhead compensation methods \cite{malony2004}.
To this end, these algorithms are specifically designed to avoid relying on information that would necessitate costly instrumentation, such as the precise tracking of memory reads and writes.
Additionally, the algorithms are intended to execute after the program has completed and, as input, take a log of all OpenMP target events.
Each event log entry must contain the start and end time of the event, the hash of the data transferred (if applicable), and the information provided by the corresponding OMPT callback, such as source and destination device numbers, code pointers, number of bytes transferred, and type of operation.

\subsection{Duplicate Data Transfers}

To detect duplicate data transfers, we employ a content-based approach using hashing.
By generating hash values of transferred data, we can identify duplicate and round-trip data transfers with minimal memory and time overhead, as only the hash values must be stored and compared.
We assume the hashing algorithm is collision-free to minimize overhead, however eliminating any chance of false positives---and thus ensuring absolute correctness---would require storing copies of every unique transfer between devices.

Algorithm \ref{alg:identify_duplicate_data_transfers} identifies duplicate data transfers by checking whether any device has received the same hash at least 2 times.
This is done efficiently by using a map to group data transfer events by their hash and receiving device number.

\begin{algorithm}
\small
\caption{Identify Duplicate Data Transfers}
\label{alg:identify_duplicate_data_transfers}
\begin{algorithmic}[1]
\Require{$data\_op\_events$ to be an array of data transfer operation events in chronological order}
\Procedure{FindDuplicateTransfers}{$data\_op\_events$}

\State let $duplicate\_transfers$ be an array$\langle$array$\langle$event$\rangle\rangle$ \LineComment{Map to lists of duplicate transfers.}
\State let $received$ be a map \\
\hspace{4.3em}$\langle \text{hash, dest\_device\_num} \rangle \to \text{array} \langle$event$\rangle$

\For{$event \in data\_op\_events$}
  \State $key \gets \langle event$.hash, $event$.dest\_device\_num$\rangle$
  \State $received$[$key$].append($event$)
\EndFor

\For{$\langle key$, $events \rangle \in received$}
  \If{$events$.size() $ < 2$}
    \State continue
  \EndIf
  \State $duplicate\_transfers$.append($events$)
\EndFor

\State\Return $duplicate\_transfers$
\EndProcedure
\end{algorithmic}
\end{algorithm}

\subsection{Round-Trip Data Transfers}

Algorithm \ref{alg:identify_round_trip_data_transfers} identifies round-trip data transfers and groups them by their hash and initial and intermediate devices.
In a similar manner to Algorithm \ref{alg:identify_duplicate_data_transfers}, we first construct a map of received data transfer events, but this time using a map to a queue, which will allow efficient extraction of events in their chronological order later.
Second, we iterate through the data transfer events, checking if a device later receives data with a matching hash, which indicates the completion of a round-trip.
To avoid falsely counting the initial transfer event as the completion trip later, we remove it from the map of received data transfer events.

\begin{algorithm}
\small
\caption{Identify Round-Trip Data Transfers}
\label{alg:identify_round_trip_data_transfers}
\begin{algorithmic}[1]
\Require{$data\_op\_events$ to be an array of data transfer operation events in chronological order}
\Procedure{FindRoundTrips}{$data\_op\_events$}

\State let $round\_trips$ be a map \\
\hspace{4.3em}$\langle \text{hash, src\_device\_num, dest\_device\_num} \rangle $\\
\hspace{4.3em}$\to $ array$\langle \text{tx\_event, rx\_event} \rangle$
\State let $received$ be a map \\
\hspace{4.3em}$\langle \text{hash, dest\_device\_num} \rangle \to \text{queue} \langle$event$\rangle$

\For{$event \in data\_op\_events$}
  \State $key \gets \langle event$.hash, $event$.dest\_device\_num$\rangle$
  \State $received$[$key$].enqueue($event$)
\EndFor

\For{$tx\_event \in data\_op\_events$}

  \State $rx\_key \gets \langle tx\_event$.hash, $tx\_event$.src\_device\_num$\rangle$
  \If{$!received$.contains($rx\_key$) or \\
  \hspace{4.1em}$received$[$rx\_key$].empty()\\
  \hspace{2.78em}}
    \LineComment{Not a round-trip, the data is never sent back.}
    \State continue 
  \EndIf

  \State $trip\_key \gets \langle tx\_event$.hash,\\
  \hspace{8.57em}$tx\_event$.src\_device\_num,\\
  \hspace{8.57em}$tx\_event$.dest\_device\_num$\rangle$
  \State $rx\_event \gets received$[$rx\_key$].peek()
  \State $round\_trips$[$trip\_key$]$.append$($\langle tx\_event$, $rx\_event \rangle$)
  \State $tx\_key \gets \langle tx\_event$.hash, $tx\_event$.dest\_device\_num$\rangle$
  \LineComment{Avoid counting this as a round-trip for other transfers.}
  \State $received$[$tx\_key$].dequeue()

\EndFor

\State\Return $round\_trips$
\EndProcedure
\end{algorithmic}
\end{algorithm}

\subsection{Repeated Device Memory Allocations}

Algorithm \ref{alg:identify_repeated_device_memory_allocations} identifies repeated device memory allocations by first pairing each allocation event with its deletion event and inserting the pair into a map where the key identifies the host and target addresses as well as the size of the allocation in bytes.
The allocation size is included as part of the key to mitigate false positives in scenarios where the same memory address is used to map different variables throughout a program's execution.
Next, we iterate through the map of allocations and remove any entries that do not have at least two allocations, leaving only repeat allocations.

\begin{algorithm}
\small
\caption{Identify Repeated Device Memory Allocations}
\label{alg:identify_repeated_device_memory_allocations}
\begin{algorithmic}[1]
\Require{$data\_op\_events$ to be an array of memory allocation/deletion events in chronological order}
\Procedure{FindRepeatedAllocs}{$data\_op\_events$}

\State let $repeated\_allocs$ be a map\\
\hspace{4.3em}$\langle \text{host\_addr, tgt\_device\_num, bytes} \rangle $\\
\hspace{4.3em}$\to $ array$\langle \text{alloc\_event, delete\_event} \rangle$

\State $allocs \gets$ \Call{GetAllocDeletePairs}{$data\_op\_events$}

\For{$\langle alloc\_event, delete\_event \rangle \in allocs$}
    \State $key \gets \langle alloc\_event$.src\_addr,\\
    \hspace{6.44em}$alloc\_event$.dest\_device\_num,\\
    \hspace{6.44em}$alloc\_event$.bytes$\rangle$
    \State $repeated\_allocs$[$key$].append($\langle alloc\_event$, $event \rangle$)
\EndFor
  
\For{$\langle key$, $events \rangle \in repeated\_allocs$}
  \If{$events$.size() $ < 2$}
    \State $repeated\_allocs$.remove($key$)
  \EndIf
\EndFor

\State\Return $repeated\_allocs$
\EndProcedure
\end{algorithmic}
\end{algorithm}

\subsection{Unused Data Mappings}

A subset of unused data mappings can be detected based on information from OpenMP event callbacks.
The idea behind Algorithms \ref{alg:identify_unused_allocs} and \ref{alg:identify_unused_transfers} is to identify data mappings that cannot possibly be made use of.
For device allocations, this means identifying all allocations whose lifetimes do not intersect with the execution of any active kernel on that device.
For data transfers, we look for any data that would be overwritten before any kernel could possibly access it or if it occurs after the last active kernel on the device.
In order to, with certainty, determine all unused data mappings we would need to know exactly what memory addresses a kernel access.
This information could be retrieved either through static analysis or intrusive memory access instrumentation, which we will investigate further in future work.

\begin{algorithm}
\small
\caption{Identify Unused Device Memory Allocation}
\label{alg:identify_unused_allocs}
\begin{algorithmic}[1]
\Require{$target\_events$ to be an array of target kernel execution events in chronological order}
\Require{$data\_op\_events$ to be an array of data transfer operation events in chronological order}
\Require{$num\_devices$ to be the number of available target devices}
\Procedure{FindUnusedAllocs}{$tgt\_events$, $data\_op\_events$}

\State let $unused\_allocs$ be an array$\langle \text{alloc\_event, delete\_event} \rangle$

\State $alloc\_events \gets$ \Call{GetAllocDeletePairs}{$data\_op\_events$}

\LineComment{Sort events by device.}
\State let $device\_tgt\_events$ be an array$\langle \text{array}\langle \text{event} \rangle\rangle$
\State let $device\_allocs$ be an array$\langle \text{array}\langle \text{event} \rangle\rangle$
\State $device\_tgt\_events \gets$ \Call{SortByDevice}{$tgt\_events$}
\State $device\_allocs\_events \gets$ \Call{SortByDevice}{$alloc\_events$}

\LineComment{Find allocations that do not overlap with target execution.}
\For{$dev\_idx $ from $0$ to $num\_devices - 1$}
  \State $\_tgt\_events \gets device\_tgt\_events$[$dev\_idx$]
  \State $\_alloc\_events \gets device\_allocs\_events$[$dev\_idx$]
  \State $tgt\_idx \gets 0$
  \For{$\langle alloc\_event, delete\_event \rangle \in \_alloc\_events$}
    \While{$tgt\_idx < \_tgt\_events$.size() and \\
    \hspace{5.5em}$\_tgt\_events$[$tgt\_idx$].end $ < alloc\_event$.start\\
    \hspace{4.3em}}
      \State $tgt\_idx \gets tgt\_idx + 1$
    \EndWhile
    \If{$tgt\_idx == \_tgt\_events$.size() or \\
        \hspace{5.5em}$\_tgt\_events$[$tgt\_idx$].start $ > delete\_event$.end\\
        \hspace{4.3em}}
    \State $unused\_allocs$.append(\\
    \hspace{13.5em}$\langle alloc\_event$, $delete\_event \rangle$)
    \EndIf 
  \EndFor

\EndFor
\State\Return $unused\_allocs$
\EndProcedure
\end{algorithmic}
\end{algorithm}

\begin{algorithm}
\small
\caption{Identify Unused Data Transfers}
\label{alg:identify_unused_transfers}
\begin{algorithmic}[1]
\Require{$target\_events$ to be an array of target kernel execution events in chronological order}
\Require{$data\_op\_events$ to be an array of data transfer operation events in chronological order}
\Require{$num\_devices$ to be the number of available target devices}
\Procedure{FindUnusedTransfers}{$tgt\_events$, $data\_op\_events$}

\State let $unused\_transfers$ be an array$\langle \text{event} \rangle$

\LineComment{Sort events by device.}
\State let $device\_tgt\_events$ be an array$\langle \text{array}\langle \text{event} \rangle\rangle$
\State let $device\_tx$ be an array$\langle \text{array}\langle \text{event} \rangle\rangle$
\State $device\_tgt\_events \gets$ \Call{SortByDevice}{$tgt\_events$}
\State $device\_tx\_events \gets$ \Call{SortByDevice}{$data\_op\_events$}

\LineComment{Find transfers from the same host address that occur more than once between target executions.}
\For{$dev\_idx $ from $0$ to $num\_devices - 1$}
  \State $\_tgt\_events \gets device\_tgt\_events$[$dev\_idx$]
  \State $\_tx\_events \gets device\_transfers\_events$[$dev\_idx$]
  \State $tgt\_idx \gets 0$
  \State let $candidates$ be a map $\langle \text{host\_addr} \rangle \to \langle \text{event} \rangle$
  \For{$tx \in \_tx\_events$}
    \While{$tgt\_idx < \_tgt\_events.size()$ and \\
    \hspace{5.5em}$\_tgt\_events$[$tgt\_idx$].end $ < tx$.start\\
    \hspace{4.3em}}
      \State $tgt\_idx \gets tgt\_idx + 1$
      \State $candidates$.clear()
    \EndWhile
    \If{$tgt\_idx == \_tgt\_events$.size()}
      \LineComment{Transfer occurs after last active kernel.}
      \State $unused\_transfers$.append($tx$)
    \ElsIf{$\_tgt\_events$[$tgt\_idx$].start $ > tx.$start}
      \LineComment{Transfer doesn't overlap with an active kernel.}
      \If{$candidates$.contains($tx$.src\_addr)}
        \State $cand \gets candidates$[$tx$.src\_addr]
        \State $unused\_transfers$.append($cand$)
      \EndIf
      \State $candidates$[$tx$.src\_addr] $ \gets tx$
    \Else
      \State $candidates$.clear()
    \EndIf
  \EndFor

\EndFor
\State\Return $unused\_transfers$
\EndProcedure
\end{algorithmic}
\end{algorithm}

Algorithms \ref{alg:identify_unused_allocs} and \ref{alg:identify_unused_transfers} are used to detect unused allocations and unused transfers, respectively.
Algorithm \ref{alg:identify_unused_allocs} first pairs each allocation event with its deletion event, then it sorts the lists of target executions and allocations by which device they occur on.
Then, for each device, it searches chronologically for any allocations whose lifetimes do not intersect with the execution of any active kernel.
Algorithm \ref{alg:identify_unused_transfers}, similarly to Algorithm \ref{alg:identify_unused_allocs}, sorts the lists of target executions and transfers by which device they occur on.
Then, to identify any transfer that is overwritten, it keeps a map of \emph{candidates}, which relates memory addresses to the last data transfer that wrote there.
If a transfer occurs and a candidate is already in the map at a common address, then the candidate was overwritten before it could be accessed and is, therefore, unused.
However, if a kernel execution occurs between transfers, then the candidates map must be cleared since the kernel may have used the data from the previous transfers.

\section{OMPDataPerf}

We developed a compiler- and hardware-agnostic performance profiling tool, \texttt{OMPDataPerf}, that integrates the algorithms described in Section \ref{sec:detection} to precisely identify the locations of issues in source code and to quantify how much optimization potential exists.

\texttt{OMPDataPerf} has two dependencies:
\begin{enumerate}
    \item The target program must be linked with an OpenMP runtime that supports the minimally compliant set of OMPT EMI callbacks.
    \item \texttt{libdw} (part of \texttt{elfutils}) must be installed on the target system.
\end{enumerate}

Specifically, \texttt{OMPDataPerf} requires the OpenMP runtime to support two OMPT callbacks: \texttt{ompt\_callback\_target\_emi} and  \texttt{ompt\_callback\_target\_data\_op\_emi}.
The OMPT callbacks required by \texttt{OMPDataPerf} are supported by most modern OpenMP runtimes implemented for compilers such as AMD-AOMP \cite{amd-aomp}, ARM-ACfL \cite{arm-acfl}, Clang \cite{llvm-clang}, and CCE \cite{hpe-cce}.
Additionally, \texttt{libdw} provides the capability to read DWARF debugging information from ELF binaries, which enhances the tool's functionality.

\begin{figure}
\centering
\includegraphics[width=\linewidth]{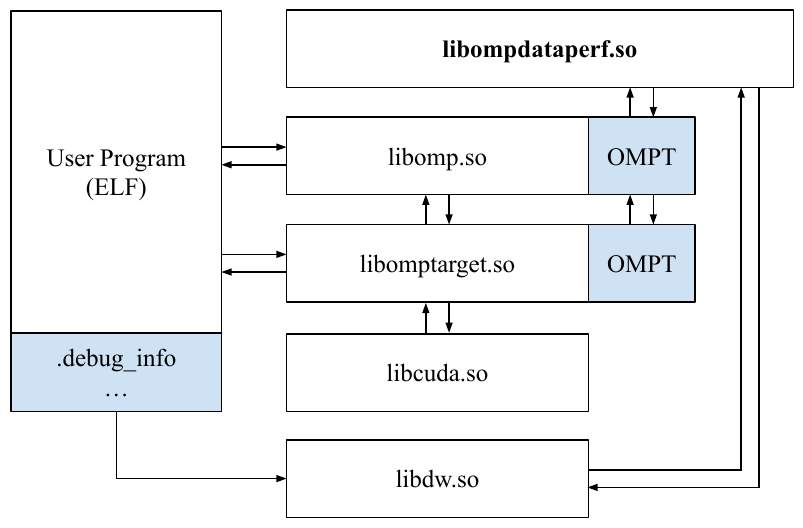}
\caption{Runtime interactions of \texttt{OMPDataPerf} with a program linked to LLVM's \texttt{libomp} OpenMP runtime offloading to an NVIDIA GPU.}
\label{fig:ompdataperf_ompt_runtime_arch}
\Description{The user's program is linked with the libomp and libomptarget OpenMP runtimes, which have OMPT interfaces. OMPDataPerf's runtime interacts with the OMPT interface during runtime to monitor execution and then uses libdw to read the debug info from the user's program.}
\end{figure}

By leveraging the OMPT interface, \texttt{OMPDataPerf} supports all OpenMP programs linked to a runtime with OMPT EMI callback compatibility. 
Programmers do \emph{not} need to recompile their code in order to use \texttt{OMPDataPerf} unless they desire line-numbers attribution; in that case the binary must be compiled with debugging information (e.g., \texttt{-g}). 
Figure \ref{fig:ompdataperf_ompt_runtime_arch} provides an overview of how \texttt{OMPDataPerf} interacts with a running program.
This specific example illustrates the integration with LLVM’s \texttt{libomp} OpenMP runtime offloading to an NVIDIA GPU.
The user's program is linked with the \texttt{libomp} and \texttt{libomptarget} runtimes (\texttt{-fopenmp}\\ \texttt{-fopenmp-targets=nvptx64}), which support the OMPT interface.

\section{Evaluation} \label{sec:evaluation}

\subsection{Experimental Setup}
Our experimental machine for measuring the runtime performance and the accuracy of its speedup prediction (Sections \ref{sec:evaluation_runtime_overhead} to \ref{sec:evaluation_accuracy_of_predicted_speedup}) was a compute node with an AMD EPYC 7543 32-core processor, \SI{512}{GB} RAM, and an NVIDIA A100-PCIE-40GB GPU running Red Hat Enterprise Linux 9.4 and CUDA 11.8.89. The benchmarks used in Section \ref{sec:evaluation_runtime_overhead} through \ref{sec:evaluation_accuracy_of_predicted_speedup} were compiled with Clang 19.1.0 and linked with the associate LLVM \texttt{libomp} OpenMP runtime of the same version. Our comparison with \texttt{Arbalest-Vec} (Section \ref{sec:evaluation_comparative_study}) was done on a compute node with an AMD EPYC 7413 24-core processor, \SI{128}{GB} RAM, and an NVIDIA A100-PCIE-80GB GPU running on Red Hat Enterprise Linux 9.4. The programs were compiled with each tool's corresponding compiler for analysis. However, for measuring the runtimes in Table \ref{tab:comparative-study-imporvements}, we compiled them with nvc++ v25.5-0.

\subsection{Benchmarks}

To evaluate \texttt{OMPDataPerf}, we selected 10 OpenMP offload benchmarks, programs, and proxy apps.
Four applications were taken from Rodinia suite \cite{rodinia-gpu}, which had OpenMP Offload versions: \texttt{bfs}, \texttt{hotspot}, \texttt{lud}, and \texttt{nw}.
\texttt{babelstream} \cite{deakin2018babelstream} was also included, a synthetic GPU memory bandwidth benchmark.
The second half of the benchmarks were selected to represent real-world HPC workloads.
This includes \texttt{minife} \cite{lin2015} from the Sandia-led Mantevo Project \cite{crozier2009, mantevo},  \texttt{minifmm} \cite{atkinson2017minifmm} from University of Bristol High Performance Computing group, \texttt{tealeaf} \cite{tealeaf-github} from the UK Mini-App Consortium and notably included in SPEChpc 2021 \cite{spechpc2021-docs}, and \texttt{rsbench} \cite{rsbench} and \texttt{xsbench} \cite{xsbench} from Argonne National Laboratory.
We selected 3 input problem sizes for each benchmark: Small, Medium, and Large.
If recommended or predefined inputs were provided we used those, otherwise inputs were selected to take roughly an order of magnitude increase in execution time when moving up input sizes.
For our comparison with \texttt{Arbalest-Vec}, we chose 5 programs from HeCBench \cite{HeCBench1} as seen from Table \ref{tab:comparative-study}.

\subsection{Runtime Overhead} \label{sec:evaluation_runtime_overhead}

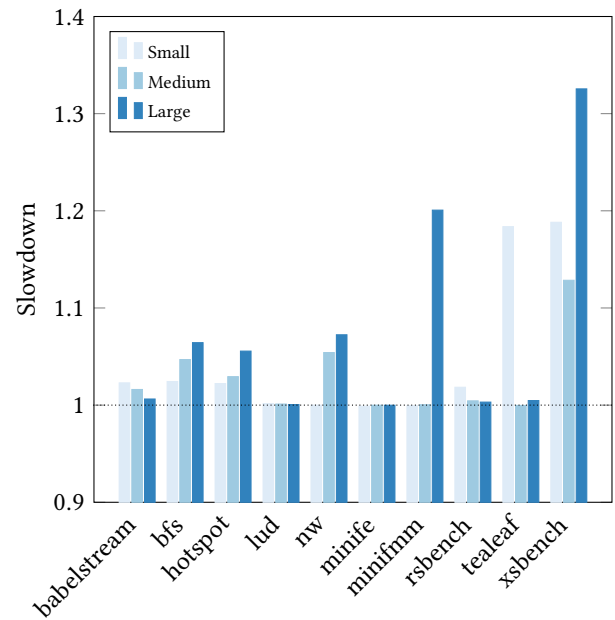
\begin{figure}[tbh]
\centering
\begin{tikzpicture}
\begin{axis}[
    width=1.0\linewidth, height=0.95\linewidth,
    ybar=2*\pgflinewidth, %
    bar width=4pt,
    enlarge x limits=0.1,
    symbolic x coords={babelstream, bfs, hotspot, lud, nw, minife, minifmm, rsbench, tealeaf, xsbench},
    xtick=data,
    ylabel={Slowdown},
    xtick style={draw=none},
    xticklabel style={rotate=45, anchor=east},
    ymin=0.9, ymax=1.4,
    legend pos=north west,
    legend style={legend columns=1, font=\scriptsize},
    legend cell align=left,
    cycle list/Blues-3,
    every axis plot post/.style={
        fill=.!100,
    },
]

\pgfplotstableread[col sep=comma]{data/profiling_overhead_apps.csv}\datatable

\addplot table[x=Program Name,y=Small]{\datatable};
\addplot table[x=Program Name,y=Medium]{\datatable};
\addplot table[x=Program Name,y=Large]{\datatable};

\addplot[densely dotted, sharp plot, update limits=false] coordinates { ([normalized]-1,1) ([normalized]10,1) };

\legend{Small, Medium, Large}
\end{axis}
\end{tikzpicture}
\caption{Runtime overhead when analyzing with \texttt{OMPDataPerf} for different input sizes. (lower is better)}
\label{fig:runtime_overhead}
\Description{Bar chart shows that OMPDataPerf has low overhead, less than 1.05 times slowdown, in most benchmarks. No discernible trend in slowdown as problem size increases.}
\end{figure}

Figure \ref{fig:runtime_overhead} shows the runtime overhead, expressed as slowdown, when using \texttt{OMPDataPerf} to analyze each program.
The largest overhead observed was a \SI{33}{\%} increase to runtime to \texttt{xsbench} on the large problem size.
Programs with more runtime dominated by host/device communication activity tended to incur greater overhead.
Seven of the ten benchmarks saw runtime overhead of less than \SI{1.07}{\times} across all problem sizes.
The geometric mean slowdown was \SI{1.05}{\times}.

\subsection{Space Overhead} \label{sec:evaluation_space_overhead}

\begin{figure}[tbh]
\centering
\begin{tikzpicture}
\begin{axis}[
    width=1.0\linewidth, height=0.95\linewidth,
    ybar=2*\pgflinewidth, %
    bar width=4pt,
    enlarge x limits=0.1,
    symbolic x coords={babelstream, bfs, hotspot, lud, nw, minife, minifmm, rsbench, tealeaf, xsbench},
    xtick=data,
    ylabel={Bytes},
    xtick style={draw=none},
    xticklabel style={rotate=45, anchor=east},
    ymode=log,
    log basis y=2, %
    legend pos=north west,
    legend style={legend columns=1, font=\scriptsize},
    legend cell align=left,
    cycle list/Blues-3,
    every axis plot post/.style={
        fill=.!100,
    },
]

\pgfplotstableread[col sep=comma]{data/space_overhead_apps.csv}\datatable

\addplot table[x=Program Name,y=Small]{\datatable};
\addplot table[x=Program Name,y=Medium]{\datatable};
\addplot table[x=Program Name,y=Large]{\datatable};

\legend{Small, Medium, Large}
\end{axis}
\end{tikzpicture}
\caption{Peak space overhead in Bytes allocated when analyzing with \texttt{OMPDataPerf} for different input sizes. (lower is better)}
\label{fig:space_overhead}
\Description{Bar chart shows that OMPDataPerf has space overhead ranging from \SI{1}{KiB} to a few \unit{MiB}.}
\end{figure}
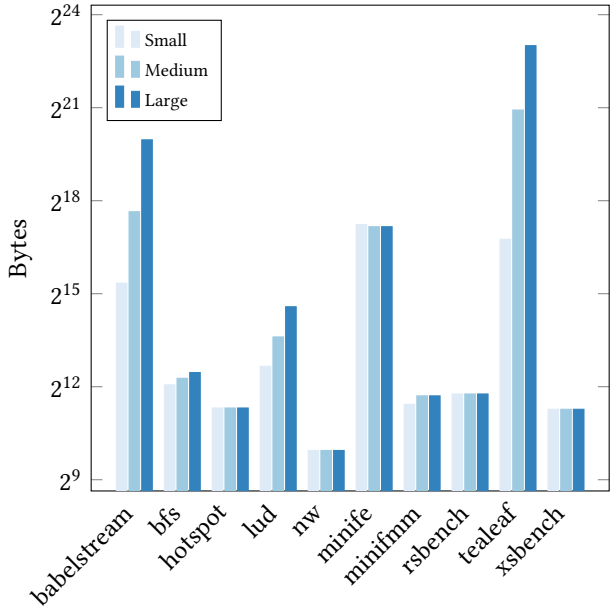

When monitoring a program, \texttt{OMPDataPerf} allocates \SI{72}{B} for every OpenMP data transfer event \SI{24}{B} for every target launch event.
The corresponding space overhead for each application is summarized in Figure \ref{fig:space_overhead} and ranges from \SI{1}{KB} to a few \unit{MB}.
The memory overhead of \texttt{OMPDataPerf} is small enough as to \emph{not} be problematic on modern HPC systems even with large benchmarks with many OpenMP target events.
\texttt{tealeaf} has by far the highest rate of overhead allocation at $\sim$\SI{1}{MB/s} (or $\sim$\SI{4}{GB/h}) of uncompressed space for event logging.
The geometric mean rate of overhead accumulation across all applications is $\sim$\SI{43}{KB/s} (or $\sim$\SI{155}{MB/h}).

\subsection{Issues Identified} \label{sec:evaluation_issues_identified}

\begin{table}
\centering
\caption{Issues Detected by \texttt{OMPDataPerf}}
\begingroup
\setlength{\tabcolsep}{2.7pt} %
\begin{tabularx}{0.8802\linewidth}{
    >{\raggedright\arraybackslash}p{0.30\linewidth}
    >{\centering\arraybackslash}p{0.09\linewidth}
    >{\centering\arraybackslash}p{0.09\linewidth}
    >{\centering\arraybackslash}p{0.09\linewidth}
    >{\centering\arraybackslash}p{0.09\linewidth}
    >{\centering\arraybackslash}p{0.09\linewidth}
}
\toprule
Program Name & DD & RT & RA & UA & UT \\ \midrule
babelstream & 499 & 0 & 499 & 0 & 0 \\
bfs & 18 & 10 & 9 & 0 & 0 \\
hotspot & 2 & 0 & 0 & 0 & 0 \\
lud & 0 & 0 & 0 & 0 & 0 \\
minife & 402 & 4 & 398 & 0 & 0 \\
minifmm & 3 & 0 & 0 & 0 & 0 \\
nw & 0 & 0 & 0 & 0 & 0 \\
rsbench & 0 & 1 & 0 & 0 & 0 \\
tealeaf & 4720 & 11 & 4706 & 0 & 0 \\
xsbench & 0 & 1 & 0 & 0 & 0 \\ \midrule
\multicolumn{6}{c}{\textit{Applications With Injected Synthetic Issues.}} \\ \midrule
babelstream (syn) & 499 & 0 & 499 & 0 & 0 \\
hotspot (syn) & 12 & 4 & 10 & 0 & 0 \\
lud (syn) & 1737 & 1243 & 747 & 250 & 252 \\
minifmm (syn) & 75 & 64 & 57 & 57 & 76 \\
nw (syn) & 8 & 0 & 4 & 1 & 3 \\
tealeaf (syn) & 17408 & 25614 & 4706 & 0 & 1 \\ \midrule
\multicolumn{6}{c}{\textit{Applications With Key Issues Fixed.}} \\ \midrule
bfs (fix) & 1 & 0 & 0 & 0 & 0 \\
minife (fix) & 3 & 0 & 0 & 0 & 0 \\
rsbench (fix) & 0 & 0 & 0 & 0 & 0 \\
xsbench (fix) & 0 & 0 & 0 & 0 & 0 \\ \bottomrule
\end{tabularx}
\endgroup
\label{tab:issues_detected}
\end{table}

To evaluate whether \texttt{OMPDataPerf} can detect data movement inefficiencies in actual programs, we perform a case study of the issues identified by the tool in each benchmark.
Based on the issues (or lack thereof) identified by \texttt{OMPDataPerf}, we either attempt to fix the issues or inject artificial ones into the program.
In the next section, Section \ref{sec:evaluation_accuracy_of_predicted_speedup}, we will evaluate whether \texttt{OMPDataPerf} correctly accesses the runtime severity of each issue.
The following abbreviations are used throughout this section: Duplicate Data Transfers (DD), Round-Trip Data Transfers (RT), Repeated Device Memory Allocations (RA), Unused Device Memory Allocations (UA), and Unused Data Transfers (UT).

Some benchmarks, including \texttt{minifmm}, \texttt{nw}, \texttt{tealeaf}, have DDs when data is first mapped on the device during initialization, e.g., multiple zero-initialized arrays of the same length.
In these cases, the DDs occur while the program is initializing the problem and not in performance-critical code, so they aren't worth fixing. \texttt{babelstream} is a GPU memory benchmark and the DDs and RAs are caused by reallocating and transferring data and results between repeated test runs, which appears to be an intentional part of the benchmark.
The \texttt{bfs} program from the Rodinia suite exhibits 3 issue types as a result of reallocating transferring back and forth a boolean to indicate when to stop launching kernels.
We eliminated these issues by moving the loop check into the OpenMP \texttt{target} region, which resulted in \SI{2.1}{\times} speedup for the small problem size.
The majority of the DDs and all of the RAs in \texttt{tealeaf} were caused by copies for initialization reduction variables.
Unfortunately, this is usually the fastest way to initialize reduction variables with current OpenMP features from our experiments.
We could not determine a performant way to eliminate these issues in \texttt{tealeaf}.
The issues detected in \texttt{minife} were fixable by extending the lifetime of intermediate variables use on the target device and result in a speedup of \SI{1.07}{\times} for the large problem size.
Both \texttt{rsbench} and \texttt{xsbench} both had a single RT caused by a missing map clause for the input struct, which unnecessarily copied the input back from the GPU; we fixed these issues.
For the benchmarks that were already well optimized, we injected artificial issues meant to mimic common inefficiencies (e.g., the issues we are trying to detect) that a programmer may stumble into around key kernels.
Table \ref{tab:issues_detected} summarizes the number of issues of each category identified by \texttt{OMPDataPerf} in each application.

\subsection{Accuracy of Predicted Optimization Potential} \label{sec:evaluation_accuracy_of_predicted_speedup}

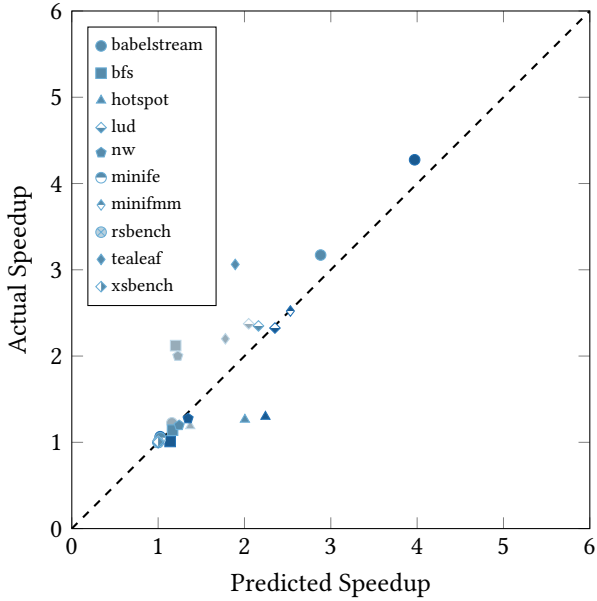
\begin{figure}
\centering
\begin{tikzpicture}

\definecolor{Blues-3-1}{HTML}{bdd7e7}
\definecolor{Blues-3-2}{HTML}{6baed6}
\definecolor{Blues-3-3}{HTML}{2171b5}
\pgfplotsset{
    discard if not/.style 2 args={
        x filter/.code={
            \edef\tempa{\thisrow{#1}}
            \edef\tempb{#2}
            \ifx\tempa\tempb
            \else
                \def\pgfmathresult{inf}
            \fi
}}}
\begin{axis}[
    xlabel={Predicted Speedup},
    ylabel={Actual Speedup},
    legend pos=north west,
    legend style={legend columns=1, font=\scriptsize},
    legend cell align={left},
    width=1.0\linewidth, height=1.0\linewidth,
    xmin=0.0, xmax=6.0,
    ymin=0.0, ymax=6.0,
    axis equal,
    cycle list name=mark list*,
]

\addplot[
    domain=\pgfkeysvalueof{/pgfplots/xmin}:\pgfkeysvalueof{/pgfplots/xmax},
    thick,
    dashed,
    color=black,
    forget plot,
] {x};

\addplot+[
    table/x=Predicted, table/y=Actual,
    only marks,
    forget plot,
        x filter/.code={
          \IfStrEq{\thisrow{Program Name}}{babelstream}{
            \IfStrEq{\thisrow{Problem Size}}{Small}{}{\def\pgfmathresult{}}
          }{\def\pgfmathresult{}}
        }, 
    Blues-3-1,
] table [col sep=comma] {data/speedup_predicted_vs_actual.csv};
\addplot+[
    table/x=Predicted, table/y=Actual,
    only marks,
    forget plot,
        x filter/.code={
          \IfStrEq{\thisrow{Program Name}}{babelstream}{
            \IfStrEq{\thisrow{Problem Size}}{Large}{}{\def\pgfmathresult{}}
          }{\def\pgfmathresult{}}
        }, 
    Blues-3-3,
] table [col sep=comma] {data/speedup_predicted_vs_actual.csv};
\addplot+[
    table/x=Predicted, table/y=Actual,
    only marks,
        x filter/.code={
          \IfStrEq{\thisrow{Program Name}}{babelstream}{
            \IfStrEq{\thisrow{Problem Size}}{Medium}{}{\def\pgfmathresult{}}
          }{\def\pgfmathresult{}}
        }, 
    Blues-3-2,
] table [col sep=comma] {data/speedup_predicted_vs_actual.csv};
\addplot+[
    table/x=Predicted, table/y=Actual,
    only marks,
    forget plot,
        x filter/.code={
          \IfStrEq{\thisrow{Program Name}}{bfs}{
            \IfStrEq{\thisrow{Problem Size}}{Small}{}{\def\pgfmathresult{}}
          }{\def\pgfmathresult{}}
        }, 
    Blues-3-1,
] table [col sep=comma] {data/speedup_predicted_vs_actual.csv};
\addplot+[
    table/x=Predicted, table/y=Actual,
    only marks,
    forget plot,
        x filter/.code={
          \IfStrEq{\thisrow{Program Name}}{bfs}{
            \IfStrEq{\thisrow{Problem Size}}{Large}{}{\def\pgfmathresult{}}
          }{\def\pgfmathresult{}}
        }, 
    Blues-3-3,
] table [col sep=comma] {data/speedup_predicted_vs_actual.csv};
\addplot+[
    table/x=Predicted, table/y=Actual,
    only marks,
        x filter/.code={
          \IfStrEq{\thisrow{Program Name}}{bfs}{
            \IfStrEq{\thisrow{Problem Size}}{Medium}{}{\def\pgfmathresult{}}
          }{\def\pgfmathresult{}}
        }, 
    Blues-3-2,
] table [col sep=comma] {data/speedup_predicted_vs_actual.csv};
\addplot+[
    table/x=Predicted, table/y=Actual,
    only marks,
    forget plot,
        x filter/.code={
          \IfStrEq{\thisrow{Program Name}}{hotspot}{
            \IfStrEq{\thisrow{Problem Size}}{Small}{}{\def\pgfmathresult{}}
          }{\def\pgfmathresult{}}
        }, 
    Blues-3-1,
] table [col sep=comma] {data/speedup_predicted_vs_actual.csv};
\addplot+[
    table/x=Predicted, table/y=Actual,
    only marks,
    forget plot,
        x filter/.code={
          \IfStrEq{\thisrow{Program Name}}{hotspot}{
            \IfStrEq{\thisrow{Problem Size}}{Large}{}{\def\pgfmathresult{}}
          }{\def\pgfmathresult{}}
        }, 
    Blues-3-3,
] table [col sep=comma] {data/speedup_predicted_vs_actual.csv};
\addplot+[
    table/x=Predicted, table/y=Actual,
    only marks,
        x filter/.code={
          \IfStrEq{\thisrow{Program Name}}{hotspot}{
            \IfStrEq{\thisrow{Problem Size}}{Medium}{}{\def\pgfmathresult{}}
          }{\def\pgfmathresult{}}
        }, 
    Blues-3-2,
] table [col sep=comma] {data/speedup_predicted_vs_actual.csv};
\addplot+[
    table/x=Predicted, table/y=Actual,
    only marks,
    forget plot,
        x filter/.code={
          \IfStrEq{\thisrow{Program Name}}{lud}{
            \IfStrEq{\thisrow{Problem Size}}{Small}{}{\def\pgfmathresult{}}
          }{\def\pgfmathresult{}}
        }, 
    Blues-3-1,
] table [col sep=comma] {data/speedup_predicted_vs_actual.csv};
\addplot+[
    table/x=Predicted, table/y=Actual,
    only marks,
    forget plot,
        x filter/.code={
          \IfStrEq{\thisrow{Program Name}}{lud}{
            \IfStrEq{\thisrow{Problem Size}}{Large}{}{\def\pgfmathresult{}}
          }{\def\pgfmathresult{}}
        }, 
    Blues-3-3,
] table [col sep=comma] {data/speedup_predicted_vs_actual.csv};
\addplot+[
    table/x=Predicted, table/y=Actual,
    only marks,
        x filter/.code={
          \IfStrEq{\thisrow{Program Name}}{lud}{
            \IfStrEq{\thisrow{Problem Size}}{Medium}{}{\def\pgfmathresult{}}
          }{\def\pgfmathresult{}}
        }, 
    Blues-3-2,
] table [col sep=comma] {data/speedup_predicted_vs_actual.csv};
\addplot+[
    table/x=Predicted, table/y=Actual,
    only marks,
    forget plot,
        x filter/.code={
          \IfStrEq{\thisrow{Program Name}}{nw}{
            \IfStrEq{\thisrow{Problem Size}}{Small}{}{\def\pgfmathresult{}}
          }{\def\pgfmathresult{}}
        }, 
    Blues-3-1,
] table [col sep=comma] {data/speedup_predicted_vs_actual.csv};
\addplot+[
    table/x=Predicted, table/y=Actual,
    only marks,
    forget plot,
        x filter/.code={
          \IfStrEq{\thisrow{Program Name}}{nw}{
            \IfStrEq{\thisrow{Problem Size}}{Large}{}{\def\pgfmathresult{}}
          }{\def\pgfmathresult{}}
        }, 
    Blues-3-3,
] table [col sep=comma] {data/speedup_predicted_vs_actual.csv};
\addplot+[
    table/x=Predicted, table/y=Actual,
    only marks,
        x filter/.code={
          \IfStrEq{\thisrow{Program Name}}{nw}{
            \IfStrEq{\thisrow{Problem Size}}{Medium}{}{\def\pgfmathresult{}}
          }{\def\pgfmathresult{}}
        }, 
    Blues-3-2,
] table [col sep=comma] {data/speedup_predicted_vs_actual.csv};
\addplot+[
    table/x=Predicted, table/y=Actual,
    only marks,
    forget plot,
        x filter/.code={
          \IfStrEq{\thisrow{Program Name}}{minife}{
            \IfStrEq{\thisrow{Problem Size}}{Small}{}{\def\pgfmathresult{}}
          }{\def\pgfmathresult{}}
        }, 
    Blues-3-1,
] table [col sep=comma] {data/speedup_predicted_vs_actual.csv};
\addplot+[
    table/x=Predicted, table/y=Actual,
    only marks,
    forget plot,
        x filter/.code={
          \IfStrEq{\thisrow{Program Name}}{minife}{
            \IfStrEq{\thisrow{Problem Size}}{Large}{}{\def\pgfmathresult{}}
          }{\def\pgfmathresult{}}
        }, 
    Blues-3-3,
] table [col sep=comma] {data/speedup_predicted_vs_actual.csv};
\addplot+[
    table/x=Predicted, table/y=Actual,
    only marks,
        x filter/.code={
          \IfStrEq{\thisrow{Program Name}}{minife}{
            \IfStrEq{\thisrow{Problem Size}}{Medium}{}{\def\pgfmathresult{}}
          }{\def\pgfmathresult{}}
        }, 
    Blues-3-2,
] table [col sep=comma] {data/speedup_predicted_vs_actual.csv};
\addplot+[
    table/x=Predicted, table/y=Actual,
    only marks,
    forget plot,
        x filter/.code={
          \IfStrEq{\thisrow{Program Name}}{minifmm}{
            \IfStrEq{\thisrow{Problem Size}}{Small}{}{\def\pgfmathresult{}}
          }{\def\pgfmathresult{}}
        }, 
    Blues-3-1,
] table [col sep=comma] {data/speedup_predicted_vs_actual.csv};
\addplot+[
    table/x=Predicted, table/y=Actual,
    only marks,
    forget plot,
        x filter/.code={
          \IfStrEq{\thisrow{Program Name}}{minifmm}{
            \IfStrEq{\thisrow{Problem Size}}{Large}{}{\def\pgfmathresult{}}
          }{\def\pgfmathresult{}}
        }, 
    Blues-3-3,
] table [col sep=comma] {data/speedup_predicted_vs_actual.csv};
\addplot+[
    table/x=Predicted, table/y=Actual,
    only marks,
        x filter/.code={
          \IfStrEq{\thisrow{Program Name}}{minifmm}{
            \IfStrEq{\thisrow{Problem Size}}{Medium}{}{\def\pgfmathresult{}}
          }{\def\pgfmathresult{}}
        }, 
    Blues-3-2,
] table [col sep=comma] {data/speedup_predicted_vs_actual.csv};
\addplot+[
    table/x=Predicted, table/y=Actual,
    only marks,
    forget plot,
        x filter/.code={
          \IfStrEq{\thisrow{Program Name}}{rsbench}{
            \IfStrEq{\thisrow{Problem Size}}{Small}{}{\def\pgfmathresult{}}
          }{\def\pgfmathresult{}}
        }, 
    Blues-3-1,
] table [col sep=comma] {data/speedup_predicted_vs_actual.csv};
\addplot+[
    table/x=Predicted, table/y=Actual,
    only marks,
    forget plot,
        x filter/.code={
          \IfStrEq{\thisrow{Program Name}}{rsbench}{
            \IfStrEq{\thisrow{Problem Size}}{Large}{}{\def\pgfmathresult{}}
          }{\def\pgfmathresult{}}
        }, 
    Blues-3-3,
] table [col sep=comma] {data/speedup_predicted_vs_actual.csv};
\addplot+[
    table/x=Predicted, table/y=Actual,
    only marks,
        x filter/.code={
          \IfStrEq{\thisrow{Program Name}}{rsbench}{
            \IfStrEq{\thisrow{Problem Size}}{Medium}{}{\def\pgfmathresult{}}
          }{\def\pgfmathresult{}}
        }, 
    Blues-3-2,
] table [col sep=comma] {data/speedup_predicted_vs_actual.csv};
\addplot+[
    table/x=Predicted, table/y=Actual,
    only marks,
    forget plot,
        x filter/.code={
          \IfStrEq{\thisrow{Program Name}}{tealeaf}{
            \IfStrEq{\thisrow{Problem Size}}{Small}{}{\def\pgfmathresult{}}
          }{\def\pgfmathresult{}}
        }, 
    Blues-3-1,
] table [col sep=comma] {data/speedup_predicted_vs_actual.csv};
\addplot+[
    table/x=Predicted, table/y=Actual,
    only marks,
    forget plot,
        x filter/.code={
          \IfStrEq{\thisrow{Program Name}}{tealeaf}{
            \IfStrEq{\thisrow{Problem Size}}{Large}{}{\def\pgfmathresult{}}
          }{\def\pgfmathresult{}}
        }, 
    Blues-3-3,
] table [col sep=comma] {data/speedup_predicted_vs_actual.csv};
\addplot+[
    table/x=Predicted, table/y=Actual,
    only marks,
        x filter/.code={
          \IfStrEq{\thisrow{Program Name}}{tealeaf}{
            \IfStrEq{\thisrow{Problem Size}}{Medium}{}{\def\pgfmathresult{}}
          }{\def\pgfmathresult{}}
        },
    Blues-3-2,
] table [col sep=comma] {data/speedup_predicted_vs_actual.csv};
\addplot+[
    table/x=Predicted, table/y=Actual,
    only marks,
    forget plot,
        x filter/.code={
          \IfStrEq{\thisrow{Program Name}}{xsbench}{
            \IfStrEq{\thisrow{Problem Size}}{Small}{}{\def\pgfmathresult{}}
          }{\def\pgfmathresult{}}
        },
    Blues-3-1,
] table [col sep=comma] {data/speedup_predicted_vs_actual.csv};
\addplot+[
    table/x=Predicted, table/y=Actual,
    only marks,
    forget plot,
        x filter/.code={
          \IfStrEq{\thisrow{Program Name}}{xsbench}{
            \IfStrEq{\thisrow{Problem Size}}{Large}{}{\def\pgfmathresult{}}
          }{\def\pgfmathresult{}}
        },
    Blues-3-3,
] table [col sep=comma] {data/speedup_predicted_vs_actual.csv};
\addplot+[
    table/x=Predicted, table/y=Actual,
    only marks,
        x filter/.code={
          \IfStrEq{\thisrow{Program Name}}{xsbench}{
            \IfStrEq{\thisrow{Problem Size}}{Medium}{}{\def\pgfmathresult{}}
          }{\def\pgfmathresult{}}
        },
    Blues-3-2,
] table [col sep=comma] {data/speedup_predicted_vs_actual.csv};

\legend{babelstream, bfs, hotspot, lud, nw, minife, minifmm, rsbench, tealeaf, xsbench}

\end{axis}
\end{tikzpicture}
\caption{Predicted Speedup vs Actual Speedup. Color intensity indicates input size (light=Small, dark=Large). Points close to the dashed line indicate that \texttt{OMPDataPerf} accurately predicted the optimization potential of the program.}
\label{fig:predicted_speedup}
\Description{Depicts a scatter plot showing predicted speedup vs actual speedup. The pointers of the scatter plot show that OMPDataProf accurately predicts the speedup for each application.}
\end{figure}

A key differentiator between \texttt{OMPDataPerf} and traditional GPU profilers, and a significant contribution of this work, is that \texttt{OMPDataPerf} can provide assessments of optimization potential, e.g. how much execution time and data management operations can be saved.
Speedup predictions are calculated by subtracting, from the total execution time, the transfer or allocation time that could be eliminated through the removal of the identified excess or inefficient data transfers and allocations.
To evaluate the accuracy of the optimization potential identified by \texttt{OMPDataPerf}, we compare the predicted speedup of the unoptimized programs to the actual speedup achieved by the optimized versions.
Figure \ref{fig:predicted_speedup} plots the predicted speedup vs the actual speedup.
We observe that \texttt{OMPDataPerf} estimates the optimization potential with an average relative error of \SI{14}{\%} and Mean Squared Error (MSE) of \SI{0.17}{\times}.
These calculations exclude one outlier: for the large problem size of \texttt{tealeaf}, the observed speedup was \SI{16}{\times}, whereas \texttt{OMPDataPerf} predicted a speedup of \SI{5.8}{\times}.
When calculating large speedups, small errors in predicted execution time can cause disproportionate errors in speedup despite the predicted time savings remaining relatively accurate and practically useful for programmers. 
Specifically, \texttt{OMPDataPerf} demonstrated \SI{90}{\%} accuracy in predicting the time savings for this case.
We believe the accuracy of \texttt{OMPDataPerf} is sufficiently high to effectively guide programmers on whether or not to pursue the optimization opportunities it identified.

\subsection{Comparison with Arbalest-Vec} \label{sec:evaluation_comparative_study}
Although Arbalest-Vec is a great tool for data correctness checking, we argue that complying with data correctness is not enough to write efficient code for heterogeneous computing. Therefore, we conduct a comparative study against Arbalest-Vec to show that \texttt{OMPDataPerf} can facilitate programmers to write efficient code.   

\texttt{Arbalest-Vec} detects the following memory anomalies: \textit{use of uninitialized memory (UUM)}, \textit{use of stale data (USD)}, \textit{use after free (UAF)}, and \textit{buffer overflow (BO)}.
Table \ref{tab:comparative-study} shows the five programs we chose from HeCBench and the reported issues from each tool. These programs were chosen because they contain kernels that are used in Computer Vision, Machine Learning, and Simulation. Here, N/A means that no issue was detected. We manually inspected and attempted to resolve the issues reported by each tool. 

Table \ref{tab:comparative-study-imporvements} shows the runtime of each program before and after fixing the identified issues. Here, \texttt{OMPDP} and \texttt{AV} refer to the measured runtime after fixing the issues reported by \texttt{OMPDataPerf} and \texttt{Arbalest-Vec}, respectively. N/A here means there was no report of issues, and FP means the detection was a False Positive, and hence no time is reported for either of the cases. Runtime was averaged over 10 trials.

\begin{table}[ht]
\centering
  \caption{Issues Detected by \texttt{OMPDataPerf} and Arbalest-Vec}
  \label{tab:comparative-study}
  \begin{tabular}{@{}lll@{}}
    \toprule
    Program Name & \texttt{OMPDataPerf} & \texttt{Arbalest-Vec}\\ \midrule
    resize-omp & DD, RA & N/A\\
    mandelbrot-omp & DD, RA, UA & UUM \\ %
    accuracy-omp & DD, UA, UT & N/A\\
    lif-omp & N/A & UUM \\ %
    bspline-vgh-omp & DD, UA, UT  & UUM \\ %
    \bottomrule
  \end{tabular}
\end{table}

\begin{table}[ht]

\centering

  \caption{Runtime Measurements Before and After Fixing the Identified Issues}
  \label{tab:comparative-study-imporvements}
  \begin{tabular}{@{}llll@{}}
    \toprule
    Program Name & Before & \texttt{OMPDP} & \texttt{AV}\\ \midrule %
    resize-omp & \SI{11.604}{s} & \SI{11.065}{s} & N/A\\ %
    mandelbrot-omp & \SI{3.974}{s} & \SI{3.950}{s} & FP \\%& 3.950s  \\
    accuracy-omp & \SI{11.644}{s} & \SI{11.640}{s} & N/A \\%& 11.640s  \\
    lif-omp & \SI{10.802}{s} & N/A & FP \\%& 10.802s\\
    bspline-vgh-omp & \SI{6.736}{s}  & \SI{5.899}{s}  & FP \\%& 5.899s\\
    \bottomrule
  \end{tabular}
\end{table}

From Table \ref{tab:comparative-study-imporvements}, \texttt{bspline-vgh} sets up to be a motivating example of why correctness checking is not sufficient to write efficient code for heterogeneous computing. As seen in Listing \ref{lst:bspline-vgh}, we removed duplicate data transfer from the original program based on \texttt{OMPDataPerf}'s suggestion. In the original program, the arrays were deterministically initialized via non-trivial multiplications of non-constant data, and they are updated at each iteration of the host-side for-loop. However, when these arrays get offloaded to the device, they are used as read-only. Thus, we fixed the redundant data movements by increasing the size of the arrays to record the initializations of each iteration, and copying the entire data just once to the device before the kernel executes. This resulted in a \SI{14}{\%} speedup in execution time, and a \SI{99}{\%} reduction in the number of calls to copy data to the device. Although this comes at the expense of taking up approximately \SI{169}{KB} more on the target device-side memory, modern GPUs can easily accommodate such extra data, and hence we claim that the modified code is more efficient. \texttt{Arbalest-Vec}, on the other hand, reported issues regarding UUMs, which we deemed to be false positives due to its conservative approach for correctness checking. Although we do not show the code for these programs due to space constraints, the reported variables were: \texttt{b[0]} and \texttt{spikes[0]} for \texttt{mandelbrot-omp} and \texttt{lif-omp}, respectively. For \texttt{bspline-vgh-omp}, it reported \texttt{walkers\_vals[0]}, \texttt{walkers\_grads[0]}, and \texttt{walkers\_hess[0]}. All of these were write-only inside the kernel.

\begin{lstlisting}[
  label=lst:bspline-vgh,
  caption={bspline-vgh-omp},
  captionpos=b,
  language=C,
  otherkeywords={define,pragma,\# },
  numbers=left,
  stepnumber=1,
  tabsize=2,
  showspaces=false,
  showstringspaces=false,
  basicstyle=\ttfamily\footnotesize,
  xleftmargin=20pt,
  float=htb,  %
  belowskip=-\parskip,
  columns=fullflexible,
  breaklines=true,
]
//before
#pragma omp target data map(alloc: a[0:4],b[0:4],c[0:4],da[0:4],db[0:4],dc[0:4],d2a[0:4],d2b[0:4], d2c[0:4]) ...{
    for (i = 0 to WSIZE-1) { ...
    //init. arrays a[0:4],b[0:4],...
    #pragma omp target update to(a[0:4])
    //update to target for remainig arrays
    ...}
}
//after
//init. arrays a[0:4*WSIZE],b[0:4*WSIZE],...
#pragma omp target data map(from:... map(to:a[0:4*WSIZE],b[0:4*WSIZE],c[0:4*WSIZE],da[0:4*WSIZE],db[0:4*WSIZE],dc[0:4*WSIZE],d2a[0:4*WSIZE],d2b[0:4*WSIZE],d2c[0:4*WSIZE]){
    for (i = 0 to WSIZE-1) {...
    //No update to target. Access each element instead
    ...}
}  
\end{lstlisting}

\subsection{Scalability}  \label{sec:scalability_and_asynchrony}

OpenMP can offload work to several GPU targets, yet most widely-used benchmarks do not expose multi-GPU support.
\texttt{OMPDataPerf} is capable of profiling programs that use multiple GPUs, but we did not evaluate this scenario because few multi‑GPU OpenMP benchmarks have been published.
For programs utilizing the asynchronous memory mapping features introduced in OpenMP 5.1 \cite{openmpSpec51} or that perform offloading to multiple GPUs, the estimates of optimization potential produced by \texttt{OMPDataPerf} may be unreliable since individual kernel or memory transfers may not necessarily translate into a proportional reduction in the overall execution time.
It should be noted that this limitation does not stem from the detection techniques employed in our approach but rather from the need to incorporate causal profiling methodologies \cite{curtsinger2015coz} within \texttt{OMPDataPerf} which can be accomplished without the need for additional instrumentation or adjustment to the algorithms presented in Section \ref{sec:detection}.
The evaluation of causal profiling for this application remains an area for future research.
\emph{Causal profiling} \cite{curtsinger2015coz} can measure optimization potential for serial, parallel, and asynchronous functions in a program without the need for instrumentation by performing timing experiments on blocks of code.

\section{Related Work} \label{sec:related_work}

Arbalest-Vec \cite{Arbalest-Vec} is a state-of-the-art dynamic analysis tool that can detect data mapping correctness issues in heterogeneous OpenMP programs.
It uses binary instrumentation in combination with the OMPT interface to achieve a high level of precision at the cost of performance, incurring an average slowdown of \SI{3.5}{\times} over native execution \cite{Arbalest-Vec}.

\texttt{OMPDataPerf}, in contrast, is a low-overhead performance profiling and analysis tool meant to guide programmer optimization efforts.
The data mapping issues diagnosed by \texttt{OMPDataPerf} are problematic for the performance of an application and not necessarily correctness.
For these reasons, we believe that \texttt{OMPDataPerf} complements Arbalest-Vec and we recommend that programmers use \texttt{OMPDataPerf} to guide their optimization efforts and Arbalest-Vec to check that they have not introduced any bugs.

In \cite{welton2018}, Welton and Miller present the discovery of several performance issues in well-known heterogeneous scientific applications.
Duplicate data transfers were found to be a performance issue in two of the applications they studied and were identified via a combination of source code review and manual instrumentation.
Later, Welton and Miller produced a prototype tool, Diogenes \cite{welton2019}, aimed at automating the identification of synchronization issues and duplicate data transfers in CUDA programs.
Diogenes profiles applications over multiple runs and uses instrumentation that results in overhead between \SI{8}{\times} and \SI{20}{\times} \cite{welton2019}, significantly higher than other performance tools.

There are many available performance tools for tracing and visualizing data transfers that can be helpful for programmers trying to find and track down data transfer performance issues in their codes.
This includes tools like NVIDIA NSight System \cite{nsys}, HPCToolkit \cite{adhianto2010hpctoolkit}, Snoopie \cite{issa2024}, and ComScribe \cite{akhtar2020comscribe}.
Unlike these tools, \texttt{OMPDataPerf} can diagnose data mapping issues without user intervention and analysis; however, \texttt{OMPDataPerf} does not currently provide visualizations of detected issues.

DrGPUM \cite{DrGPUM} and ValueExpert \cite{ValueExpert}, like \texttt{OMPDataPerf}, target GPU‑memory inefficiencies but for CUDA programs instead of OpenMP programs.
DrGPUM focuses on allocation-level problems, such as early allocation, late deallocation, and unused allocations, whereas ValueExpert targets performance issues arising from value patterns, notably redundant writes and ineffective memory updates.
In contrast, \texttt{OMPDataPerf} performs a coarser‑grained analysis that is specifically aware of OpenMP target directives.
By tracking the lifetime and content of mapped variables across successive target regions, \texttt{OMPDataPerf} can detect duplicate data transfers, round‑trip transfers, and repeated device‑allocation patterns; issues that would go undetected by both DrGPUM and ValueExpert.
While ValueExpert’s detection of redundant writes and ineffective updates is largely disjoint from \texttt{OMPDataPerf}’s detection capabilities, \texttt{OMPDataPerf} and DrGPUM have some overlap in the types of redundant and unused memory allocation issues that may be detected. 
Moreover, the content-aware approach of \texttt{OMPDataPerf} incurs a low runtime overhead of \SI{5}{\%} slowdown, whereas the median slowdowns reported for DrGPUM and ValueExpert are \SI{5.43}{\times} and \SI{7.81}{\times}, respectively, with worst-case penalties exceeding $20$ to \SI{70}{\times}.
Additionally, both DrGPUM and ValueExpert rely on extensive instrumentation, with callbacks at every memory access. This heavy instrumentation has massive overhead and prevents the extraction of meaningful timing measurements. Our techniques are low-overhead, which is paramount for profiling since it allows preservation of accurate timing information and allows programmers to more intelligently dedicate optimization effort.
We did not directly evaluate our tool against DrGPUM or ValueExpert, as they are designed for CUDA programs, whereas our tool is designed for OpenMP GPU programs.
A meaningful comparison would require workloads implemented in both programming models and that exhibit identical GPU memory-transfer behaviors.
While benchmark suites such as HeCBench provide dual CUDA/OpenMP variants, it is not guaranteed that equivalent data-movement bugs or performance-critical patterns exist in both variants, making a fair evaluation unfeasible without substantial additional engineering effort.

\section{Conclusion} \label{sec:conclusion}

In this work, we propose and implement dynamic analysis techniques into \texttt{OMPDataPerf}, a tool for profiling and detecting inefficient data transfer patterns in heterogeneous OpenMP applications.
\texttt{OMPDataPerf} was able to detect and accurately quantify the optimization potential of data mapping performance issues in prominent real-world applications, resulting in speedups up to \SI{2.1}{\times}.
Furthermore, this is accomplished with overhead averaging just \SI{5}{\%} with the worst overhead measured at \SI{33}{\%}.
By automating the detection of inefficient data mapping patterns, our tool saves programmers significant time and effort, accelerating the development of heterogeneous applications.

\begin{acks}
This research was supported by the National Science Foundation under grant numbers 2426580 and 2211982.
We would also like to thank the Research IT team\footnote{\url{https://researchit.las.iastate.edu/}} of Iowa State University for their support and for providing access to the HPC Cluster, which were essential for conducting the experiments in this project.
\end{acks}

\bibliographystyle{ACM-Reference-Format}
\bibliography{references}\balance
\clearpage
\appendix
\section{Artifact Appendix}

\subsection{Abstract}

This artifact provides the source code and benchmark applications for \texttt{OMPDataPerf}, a portable, low-overhead performance analysis tool for detecting, profiling, and attributing inefficient data-mapping patterns in OpenMP programs. 
The artifact includes the implementation of Algorithms \ref{alg:identify_duplicate_data_transfers} through \ref{alg:identify_unused_transfers}, allowing users to exercise and evaluate the tool's dynamic analysis capabilities on the provided benchmarks and their own applications.
The artifact is open-source and available at \url{https://zenodo.org/records/18102356} and \url{https://github.com/lmarzen/ompdataperf}.

\subsection{Artifact check-list (meta-information)}

{\small
\begin{itemize}
  \item {\bf Program: } \texttt{ompdataperf}
  \item {\bf Compilation: } Clang $\geq$ 19.1 (w/ OpenMP offloading support)
  \item {\bf Run‑time environment: } Linux with OpenMP 5.1 support, GPU toolchain (e.g. CUDA, ROCm) 
  \item {\bf Hardware: } x86\_64 CPU, AMD or NVIDIA GPU.
  \item {\bf Output: } Human‑readable tables.
  \item {\bf Experiments: } \texttt{babelstream}, \texttt{bfs}, \texttt{hotspot}, \texttt{lud}, \texttt{minife}, \texttt{minifmm}, \texttt{nw}, \texttt{rsbench}, \texttt{tealeaf}, \texttt{xsbench}, and custom synthetic benchmarks.
  \item {\bf How much disk space required (approximately)?: } \SI{2}{GB}
  \item {\bf How much time is needed to prepare workflow (approximately)?: } \SI{15}{minutes}.
  \item {\bf How much time is needed to complete experiments (approximately)?: } \SI{30}{minutes}.
  \item {\bf Publicly available?: } Yes
  \item {\bf Code licenses (if publicly available)?: } GNU GPL-3.0
\end{itemize}
}

\subsection{Description}

\subsubsection{How to access}\label{sec:ae-how-to-access}

The artifact is archived on Zenodo. Alternatively, it may be obtained via the public git repository.

{\footnotesize
\begin{lstlisting}[breaklines=true, basicstyle=\ttfamily, columns=fixed]
$ wget "https://zenodo.org/records/18102356/files/ompdataperf-0.1.0+submodules.tar.gz"
$ tar -xzf ompdataperf-0.1.0+submodules.tar.gz
$ cd ompdataperf
\end{lstlisting}
}

or

{\footnotesize
\begin{lstlisting}[breaklines=true, basicstyle=\ttfamily, columns=fixed]
$ git clone https://github.com/lmarzen/ompdataperf.git v0.1.0
$ cd ompdataperf
$ git submodule update --init
\end{lstlisting}
}

\subsubsection{Hardware dependencies}
{\small
\begin{itemize}
  \item x86\_64 CPU
  \item An OpenMP‑capable GPU (AMD or NVIDIA)
  \item At least \SI{2}{GB} of free RAM and \SI{2}{GB} of free disk space for code and/or the container image.
\end{itemize}
}

\subsubsection{Software dependencies}

{\small
\begin{enumerate}
  \item Clang $\ge$ 19.1 (w/ OpenMP offloading support)
  \item elfutils (libdw) $\ge$ 0.189
  \item CMake $\ge$ 3.12
  \item Python~3
  \item GPU toolchain(s):
    \begin{itemize}
      \item NVIDIA: CUDA toolkit
      \item AMD: ROCm device libraries
    \end{itemize}
\end{enumerate}
}

The provided Docker containers can (optionally) be used to simplify environment setup; see Section \ref{sec:ae-docker} for setup instructions.

\subsubsection{Data sets}\label{sec:ae-data-sets}

The Rodinia 3.1 datasets are used as input to some benchmarks. They are not included in the artifact and should be downloaded separately.
{\footnotesize
\begin{lstlisting}[breaklines=true, basicstyle=\ttfamily, columns=fixed]
$ bash eval/download_dataset.sh
\end{lstlisting}
}

\subsection{Installation}

\subsubsection{Docker (optional)}\label{sec:ae-docker}

\texttt{OMPDataPerf} supports two \\Dockerfiles, one for NVIDIA CUDA and the other for AMD ROCM. Reviewers should select the docker container corresponding to the GPU available on their system.
\\
\noindent\textit{NVIDIA GPUs}:

To get GPU support for NVIDIA GPUs working in a docker container you may require additional setup on your host system:
{\footnotesize
\begin{lstlisting}[breaklines=true, basicstyle=\ttfamily, columns=fixed]
$ sudo apt-get install -y nvidia-container-toolkit
$ sudo nvidia-ctk runtime configure --runtime=docker
$ sudo systemctl restart docker
\end{lstlisting}
}

\noindent To run the container for Nvidia GPUs:
{\footnotesize
\begin{lstlisting}[breaklines=true, basicstyle=\ttfamily, columns=fixed]
$ cd docker/cuda-docker-image
$ docker build -t ompdataperf:latest .
$ docker run --rm -it --gpus all ompdataperf:latest bash
\end{lstlisting}
}

\noindent\textit{AMD GPUs}:

\noindent To run the container for AMD GPUs:
{\footnotesize
\begin{lstlisting}[breaklines=true, basicstyle=\ttfamily, columns=fixed]
$ cd docker/rocm-docker-image
$ docker build -t ompdataperf:latest .
$ docker run --rm -it --device=/dev/kfd --device=/dev/dri --group-add video ompdataperf:latest bash
\end{lstlisting}
}

After the docker environment is setup, proceed to Section \ref{sec:ae-compiling} to build \texttt{OMPDataPerf}.

\subsubsection{Compiling \texttt{OMPDataPerf}}\label{sec:ae-compiling}

Once your environment is prepared, you should download and extract the artifact, see Section \ref{sec:ae-how-to-access}. Then run the \texttt{build.sh} bash script to invoke CMake and compile \texttt{OMPDataPerf}.
{\footnotesize
\begin{lstlisting}[breaklines=true, basicstyle=\ttfamily, columns=fixed]
$ bash build.sh
\end{lstlisting}
}

\subsection{Experiment workflow}

This section walks through the steps necessary to analyze the benchmark applications used in the evaluations in Section \ref{sec:evaluation} using \texttt{OMPDataPerf}.

\subsubsection{Data sets} See Section \ref{sec:ae-data-sets}, for instructions on how to download the Rodinia data sets used as input to some benchmarks.

\subsubsection{Build Evaluation Programs}
The build scripts for the included benchmarks were intended for use on a machine with an Nvidia GPU, if you have an AMD GPU you’ll need to replace the -fopenmp-targets=nvptx flag. This can be quickly done with the following command. (for AMD GPUs only):
{\footnotesize
\begin{lstlisting}[breaklines=true, basicstyle=\ttfamily, columns=fixed]
$ find . -type f -exec sed -i 's/nvptx64-nvidia-cuda/amdgcn-amd-amdhsa/g; s/nvptx64/amdgcn-amd-amdhsa/g' {} +
\end{lstlisting}
}
\noindent Compile the benchmarks used for evaluation:
{\footnotesize
\begin{lstlisting}[breaklines=true, basicstyle=\ttfamily, columns=fixed]
$ cd eval
$ make
\end{lstlisting}
}

\subsubsection{Usage}\label{sec:ae-usage}\hspace{0pt}
{\footnotesize
\begin{lstlisting}[breaklines=true, basicstyle=\ttfamily, columns=fixed]
Usage: ompdataperf [options] [program] [program arguments]
Options:
  -h, --help      Show this help message
  -q, --quiet     Suppress warnings
  -v, --verbose   Enable verbose output
  --version       Print the version of ompdataperf
\end{lstlisting}
}

\noindent Profile \texttt{hotspot} with \texttt{OMPDataPerf}:
{\footnotesize
\begin{lstlisting}[breaklines=true, basicstyle=\ttfamily, columns=fixed]
$ ../build/ompdataperf ./src/hotspot/hotspot_offload 64 64 2 4 data/hotspot/temp_64 data/hotspot/power_64 output.out
\end{lstlisting}
}

\subsection{Evaluation and expected results}

Running the command in Section \ref{sec:ae-usage} should produce output similar to the excerpt below.  Minor timing variations are expected.

{\tiny
\begin{lstlisting}[breaklines=true, basicstyle=\ttfamily, columns=fixed]
info: OpenMP OMPT interface version TR4 5.0 preview 1
info: OpenMP runtime LLVM OMP version: 5.0.20140926
warning: OMPDataPerf requires OMPT interface version 5.1 (or later), but found version TR4 5.0 preview 1. Some features may be degraded.
Start computing the transient temperature
Ending simulation
Total time: 0.010089 seconds

=== OpenMP Duplicate Target Data Transfer Analysis ===
   time(%
     0.11%
     0.11%

=== OpenMP Round-Trip Target Data Transfer Analysis ===
...
\end{lstlisting}
}

\subsection{Experiment customization}

This artifact is \textit{reusable} and we encourage its use to analyze programs beyond the applications used in our evaluation. The expirments can be further customized by using alternative hardware and compilers. \texttt{OMPDataPerf} is hardware- and compiler-agnostic and is functional so long as the environment has an OpenMP 5.1 compliant runtime that supports GPU offloading to the hardware.

\subsubsection{Demonstration of Reusability}
HeCBench is a repository that provides a collection of hetergeneous computing benchmarks, most of which were not used in our evaluation.

{\footnotesize
\begin{lstlisting}[breaklines=true, basicstyle=\ttfamily, columns=fixed]
$ git clone https://github.com/zjin-lcf/HeCBench.git
\end{lstlisting}
}

\texttt{accuracy} is a simple benchmark in HeCBench that you can use for the purpose of validating that this artifact is reusable. Of course, you can choose any OpenMP GPU program of your choosing.
{\footnotesize
\begin{lstlisting}[breaklines=true, basicstyle=\ttfamily, columns=fixed]
$ cd HeCBench/src/accuracy-omp
$ make -f Makefile.aomp    # for AMD GPUs
$ make -f Makefile.nvc     # for NVIDIA GPUs
\end{lstlisting}
}
\noindent Analyze with OMPDataPerf.
{\footnotesize
\begin{lstlisting}[breaklines=true, basicstyle=\ttfamily, columns=fixed]
$ ./path/to/build/ompdataperf ./main 8192 10000 10 100
\end{lstlisting}
}

\subsection{Methodology}

Submission, reviewing and badging methodology:

\begin{itemize}
  \item \url{https://www.acm.org/publications/policies/artifact-review-and-badging-current}
  \item \url{https://cTuning.org/ae}
\end{itemize}

\section{Hash}
\subsection{Hash Selection} \label{sec:evaluation_hash_selection}
Selecting a high-quality and performant hash function for this use-case is important else users might encounter confusing false positives or experience significant runtime overhead.
We evaluated 19 non-cryptographic hashes to identify the most performant hashes of sufficient quality in this application.
The hashes we evaluated were selected from 6 hash function families that passed most or all of the quality tests in the SMHasher3 suite \cite{smhasher3-gitlab}.
CityHash \cite{cityhash-github} and it's successor FarmHash \cite{farmhash-github} are hash functions developed by Google for x86 32 and 64-bit CPUs.
MeowHash \cite{meowhash-github} is optimized for hashing long strings on x86 64-bit CPUs with hardware AES acceleration.
rapidhash \cite{rapidhash-github} is the official successor to wyhash \cite{wyhash-github}, the default hashing algorithm in the Go and Zig languages, optimized for AMD64 and AArch64 CPUs.
t1ha \cite{t1ha-github} was written for x86 64-bit CPUs and uses SIMD-friendly algorithms.
xxHash \cite{xxhash-github} is very portable and is optimized for 32 and 64-bit CPUs, and its newer XXH3 hashes support SIMD.

\begin{table*}
\caption{Hash Rate in \unit{GB/s} for Medium Problem Sizes}
\centering
\begingroup
\pgfplotstableset{
    typeset cell/.append code={%
        \ifnum\pgfplotstablerow<0%
            \ifnum\pgfplotstablecol=\pgfplotstablecols%
                \pgfkeyssetvalue{/pgfplots/table/@cell content}{\rotatebox{90}{#1}\\}%
            \else
                \ifnum\pgfplotstablecol>1%
                    \pgfkeyssetvalue{/pgfplots/table/@cell content}{\rotatebox{90}{#1}&}%
                \fi
            \fi
        \fi
    },
    /color cells/min/.initial=0,
    /color cells/max/.initial=100,
    /color cells/textcolor/.initial=,
    color cells/.code={%
        \pgfqkeys{/color cells}{#1}%
        \pgfkeysalso{%
            postproc cell content/.code={%
                \begingroup
                \pgfkeysgetvalue{/pgfplots/table/@preprocessed cell content}\value
                \ifx\value\empty
                    \endgroup
                \else
                \pgfmathfloatparsenumber{\value}%
                \pgfmathfloattofixed{\pgfmathresult}%
                \let\value=\pgfmathresult
                \pgfplotscolormapaccess
                    [\pgfkeysvalueof{/color cells/min}:\pgfkeysvalueof{/color cells/max}]%
                    {\value}%
                    {\pgfkeysvalueof{/pgfplots/colormap name}}%
                \pgfkeysgetvalue{/pgfplots/table/@cell content}\typesetvalue
                \pgfkeysgetvalue{/color cells/textcolor}\textcolorvalue
                \toks0=\expandafter{\typesetvalue}%
                \xdef\temp{%
                    \noexpand\pgfkeysalso{%
                        @cell content={%
                            \noexpand\rule{0cm}{2.4ex}%
                            \noexpand\cellcolor[rgb]{\pgfmathresult}%
                            \noexpand\definecolor{mapped color}{rgb}{\pgfmathresult}%
                            \ifx\textcolorvalue\empty
                            \else
                                \noexpand\color{\textcolorvalue}%
                            \fi
                            \the\toks0 %
                        }%
                    }%
                }%
                \endgroup
                \temp
                \fi
            }%
        }%
    },
    every column/.style={
        column type=>{\centering\arraybackslash}p{6.87mm}
    },
    every first column/.style={
        reset styles,
        string type,
        column type=r
    },
    every head row/.style={
        column type=c
    },
}
\makeatletter\@makeother\_
\setlength{\tabcolsep}{2pt} %
\begin{filecontents*}{data/hash_rate_apps.csv}
Program Name,CityHash32,CityHash64,CityHash128,CityHashCrc128,FarmHash32,FarmHash64,FarmHash128,MeowHash,rapidhash,t1ha0_avx,t1ha0_avx2,t1ha0_noavx,t1ha0_32le,t1ha1_le,t1ha2_atonce,XXH32,XXH64,XXH3_64bits,XXH3_128bits
babelstream,4.5,15.0,15.6,17.8,23.8,23.8,15.7,28.3,22.4,26.9,28.9,27.2,8.4,15.7,15.6,6.9,9.9,27.8,26.1
bfs,4.3,14.5,14.9,17.0,26.4,26.2,15.0,37.1,22.7,28.7,48.0,29.0,8.3,15.3,15.4,6.8,9.9,41.1,40.2
hotspot,3.7,11.7,12.8,14.2,21.2,20.8,12.6,27.4,18.0,23.6,32.2,23.5,6.9,12.5,13.1,6.2,8.8,29.0,29.7
lud,4.5,15.2,15.8,18.2,24.3,24.4,15.9,33.4,23.1,28.1,30.4,27.4,8.5,15.9,15.8,7.0,10.0,28.2,27.8
minife,4.2,14.1,14.6,17.4,23.6,23.6,14.5,31.5,22.5,27.9,29.2,27.4,7.8,14.4,14.1,6.2,9.4,27.1,26.6
minifmm,3.6,11.7,11.6,14.2,15.8,16.5,11.6,33.5,21.2,25.8,40.1,25.4,7.0,11.9,12.1,5.4,8.2,25.7,25.1
nw,4.5,14.9,15.5,17.8,24.4,24.2,15.5,32.6,22.0,27.3,34.8,26.8,8.5,15.6,15.6,7.0,10.0,31.3,30.7
rsbench,4.5,14.7,15.4,17.0,21.8,21.8,15.4,26.8,20.1,23.7,28.8,23.8,8.6,15.4,15.1,7.0,10.1,26.1,25.5
tealeaf,3.5,9.5,8.5,9.1,11.8,12.2,8.0,13.5,11.6,13.2,16.9,14.1,6.0,9.6,9.3,4.9,6.6,16.0,15.3
xsbench,4.5,15.1,15.8,18.3,24.2,24.2,15.8,32.5,23.2,28.9,30.7,28.5,8.4,15.9,15.6,6.9,9.9,27.9,27.6
AVERAGE,4.2,13.6,14.0,16.1,21.7,21.8,14.0,29.6,20.7,25.4,32.0,25.3,7.8,14.2,14.2,6.4,9.3,28.0,27.5
\end{filecontents*}
\pgfplotstabletypeset[%
    color cells={min=3.5,max=48.0},
    /pgfplots/colormap={blackwhite}{color=(white); rgb255=(8,81,156)},
    /pgf/number format/fixed zerofill,
    /pgf/number format/precision=1,
    col sep=comma
]{data/hash_rate_apps.csv}
\endgroup
\label{tab:hash_rates_apps}
\end{table*}

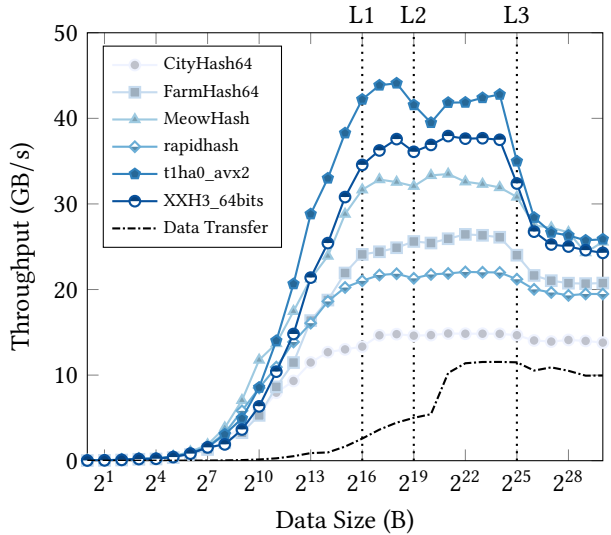
\begin{figure}[b]
\centering
\begin{tikzpicture}
\begin{axis}
[
    xlabel={Data Size (\unit{B})}, 
    ylabel={Throughput (\unit{GB/s})},
    xmin=1, xmax=2^30, 
    ymin=0, ymax=50,
    xmode=log,
    log basis x=2, %
    xtick distance=2^3,
    clip mode=individual, %
    legend pos=north west,
    legend style={legend columns=1, font=\scriptsize},
    legend cell align={left},
    cycle list/Blues-6,
    cycle multiindex* list={
        mark list*\nextlist
        Blues-6\nextlist
    },
    every axis plot post/.style={
        thick,
    },
]

\addplot[dotted, thick, black, forget plot] coordinates {(2^16,\pgfkeysvalueof{/pgfplots/ymin}) (2^16,\pgfkeysvalueof{/pgfplots/ymax})};
\addplot[dotted, thick, black, forget plot] coordinates {(2^19,\pgfkeysvalueof{/pgfplots/ymin}) (2^19,\pgfkeysvalueof{/pgfplots/ymax})};
\addplot[dotted, thick, black, forget plot] coordinates {(2^25,\pgfkeysvalueof{/pgfplots/ymin}) (2^25,\pgfkeysvalueof{/pgfplots/ymax})};

\node[anchor=south] at (axis cs:2^16,\pgfkeysvalueof{/pgfplots/ymax}) {L1};
\node[anchor=south] at (axis cs:2^19,\pgfkeysvalueof{/pgfplots/ymax}) {L2};
\node[anchor=south] at (axis cs:2^25,\pgfkeysvalueof{/pgfplots/ymax}) {L3};

\addplot table[col sep=comma, x=Input Size, y=CityHash64] {data/hash_rate_torture.csv};
\addplot table[col sep=comma, x=Input Size, y=FarmHash64] {data/hash_rate_torture.csv};
\addplot table[col sep=comma, x=Input Size, y=MeowHash] {data/hash_rate_torture.csv};
\addplot table[col sep=comma, x=Input Size, y=rapidhash] {data/hash_rate_torture.csv};
\addplot table[col sep=comma, x=Input Size, y=t1ha0_ia32aes_avx2] {data/hash_rate_torture.csv};
\addplot table[col sep=comma, x=Input Size, y=XXH3_64bits] {data/hash_rate_torture.csv};

\addlegendentry{CityHash64}
\addlegendentry{FarmHash64}
\addlegendentry{MeowHash}
\addlegendentry{rapidhash}
\addlegendentry{t1ha0\_avx2}
\addlegendentry{XXH3\_64bits}

\addplot[thick, densely dashdotted, black] table[col sep=comma, x=Input Size, y=Rate] {data/transfer_rate_torture.csv};
\addlegendentry{Data Transfer}

\end{axis}
\end{tikzpicture}
\caption{Average sequential throughput vs data size. (higher is better)}
\label{fig:hash_rate_vs_data_size}
\Description{Graph shows that sequential hashing has higher throughput and lower start-up cost when compared to sequential data transfer to a GPU at all input sizes. Sequential hashing throughput peaks before exceeding the capacity of a core's L2 cache and sees a larger drop when exceeding the L3 cache.}
\end{figure}

To enable the evaluation of hash quality for this use case, we added an optional feature to \texttt{OMPDataPerf} that stores copies of all transferred data and checks for hash collisions.
While this feature incurs moderate runtime overhead and extremely high memory overhead, it allows comprehensive collision detection when enabled.
Due to the extreme overhead, this option is disabled by default but, when used, allows us to check for collisions.
Across all benchmarks and problem sizes, we observed $0$ collisions for all evaluated hash functions.

To evaluate the performance of the selected hash functions under real-world conditions, we instrumented \texttt{OMPDataPerf} with a timer to measure the effective hash rate.
We found that \texttt{t1ha0\_avx2} has the fastest average hash rate (\SI{32}{GB/s}) for medium and large problem sizes, while \texttt{MeowHash} had the fastest average hash rate (\SI{25}{GB/s}) for small problem sizes.
Table \ref{tab:hash_rates_apps} shows the effective hash rate of the data transferred for each hash function in \unit{GB/s} for each benchmark with medium problem size.
We selected \texttt{t1ha0\_avx2} to be the default hash function for \texttt{OMPDataPerf} since it consistently performed well across all problem sizes with an average finishing position of $1.3$.

Figure \ref{fig:hash_rate_vs_data_size} compares the effective hash rate of the top-\\performing hash functions from each hash family as a function of data size.
These results were obtained using a synthetic memory transfer torture test we developed for this purpose.
The fastest non-cryptographic hash functions are faster than sequential reads from main memory.
As a consequence of this, we observe a significant decrease in hash rate when the size of contiguous data exceeds the capacity of the \SI{32}{MiB} L3 cache shared among the cores in each Core Complex (CCX) of the AMD EPYC 7543 processor.
We also plotted the throughput of host/device data transfers, which reaffirms that data transfers have higher startup costs and require substantially larger data volumes to achieve peak throughput.
The top-performing hash functions demonstrated higher effective throughput than host/device data transfers for a single sequential data stream.
This is encouraging since the hashing operation is the most computationally expensive runtime operation we rely on to implement the detection methods we presented.
Small data volumes, \SI{64}{B} or less (e.g., primitives and structs), can be hashed $100$ to \SI{200}{\times} faster than they can be transferred.
Once the data size exceeds the L3 cache and the hash rate decreases, hashing is still $2.4$ to \SI{3.0}{\times} the rate of data transfers, fast enough that the runtime impact should remain manageable.

\begin{table*}
\centering
\caption{Programs and Inputs Used for Evaluating {\tt OMPDataPerf}}
\begingroup
\renewcommand{\arraystretch}{1.15} %
\setlength{\tabcolsep}{2pt} %
\begin{tabularx}{\linewidth}{
    >{\raggedright\arraybackslash}p{0.14\linewidth}
    >{\centering\arraybackslash}p{0.17\linewidth}
    >{\centering\arraybackslash}p{0.217\linewidth}
    >{\centering\arraybackslash}p{0.217\linewidth}
    >{\centering\arraybackslash}p{0.217\linewidth}
}
\toprule
\multicolumn{1}{c}{\multirow{2}{*}{Application}} & \multirow{2}{*}{Domain} & \multicolumn{3}{c}{Input/Problem Size} \\
\multicolumn{1}{c}{} &  & Small & Medium & Large \\ \midrule
babelstream \cite{deakin2018babelstream} & Memory Bandwidth & -n 100 -s 1048576 & -n 500 -s 33554432 & -n 2500 -s 33554432 \\
bfs \cite{rodinia-gpu} & Graph Algorithms & graph4096.txt & graph65536.txt & graph1MW\_6.txt \\
hotspot \cite{rodinia-gpu} & Thermal Simulation & {\footnotesize 64 64 2 4 temp\_64 power\_64} & {\footnotesize 512 512 2 4 temp\_512 power\_512} & {\footnotesize 1024 1024 2 4 temp\_1024 power\_1024} \\
lud \cite{rodinia-gpu} & Linear Algebra & -s 2000 & -s 4000 & -s 8000 \\
minife \cite{mantevo} & Finite Element Analysis & -nx 66 -ny 64 -nz 64 & -nx 132 -ny 128 -nz 128 & -nx 264 -ny 256 -nz 256 \\
minifmm \cite{atkinson2017minifmm} & Particle Physics & -n 100 & -n 1000 & -n 10000 \\
nw \cite{rodinia-gpu}& Bioinformatics & 512 10 2 & 2048 10 2 & 8192 10 2 \\
rsbench \cite{rsbench} & Neutron Transport & -m event -s small & -m event -s large -l 4250000 & -m event -s large \\
tealeaf \cite{mcintosh-smith2017tealeaf} & High Energy Physics & --file tea\_bm\_1.in & --file tea\_bm\_2.in & --file tea\_bm\_4.in \\
xsbench \cite{xsbench} & Neutron Transport & -m event -s small & -m event -g 1413 & -m event -s large \\ \bottomrule
\end{tabularx}
\endgroup
\label{tab:benchmarks_desc}
\end{table*}

\section{Benchmark Inputs}
Table \ref{tab:benchmarks_desc} shows the inputs used for the benchmarks used in Sections 7.3 through 7.6. The inputs used for Section 7.7 were the default inputs in the Makefile of each subdirectory in the HeCBench \cite{HeCBench1}. 

\section{OMPT Compiler and Runtime Support} \label{sec:survey}

The \emph{OpenMP Tools Interface} (OMPT) and its subsequent extension with the \emph{External Monitoring Interface} (EMI) callbacks in OpenMP 5.1 has seen varying levels of adoption across different compilers and their associated runtimes.
This section provides an overview of the current compiler support for target-related OMPT features, which aims to inform future strategies for tool development for heterogeneous OpenMP applications.
We investigate and compare compilers from prominent hardware vendors, including AMD, ARM, Intel, and NVIDIA, as well as compilers from the GNU and LLVM projects. 
To assess compiler support for OpenMP OMPT features, we conducted a comprehensive review of publicly available documentation and manuals, audited source code (when available), and conducted basic testing.

\begin{table*}[htb]
\renewcommand\thempfootnote{\fnsymbol{mpfootnote}}
\begin{minipage}{\textwidth}
\centering
\caption{Compiler and Runtime Support of OMPT Target Features}
\begingroup
\setlength{\tabcolsep}{0pt} %
\begin{tabularx}{\linewidth}{
    >{\centering\arraybackslash}p{3em}
    >{\raggedright\arraybackslash}p{6.5em}
    >{\raggedright\arraybackslash}p{3.5em}
    >{\centering\arraybackslash}p{3.5em}
    >{\centering\arraybackslash}p{3.5em}
    >{\centering\arraybackslash}p{3.5em}
    >{\centering\arraybackslash}p{3.5em}
    >{\centering\arraybackslash}p{3.5em}
    >{\centering\arraybackslash}p{3.5em}
    >{\centering\arraybackslash}p{4.5em}
    >{\centering\arraybackslash}p{4.5em}
    >{\centering\arraybackslash}p{4.5em}
}
\toprule
\begin{tabular}[c]{@{}c@{}}OMP\\ Std.\end{tabular}
& \multicolumn{2}{c}{Feature Category}
& \begin{tabular}[c]{@{}c@{}}AMD \\ {\small AOCC}\end{tabular}
& \begin{tabular}[c]{@{}c@{}}AMD\\ {\small AOMP}\end{tabular}
& \begin{tabular}[c]{@{}c@{}}AMD\\ {\small ROCm}\end{tabular}
& \begin{tabular}[c]{@{}c@{}}Arm\\ {\small ACfL}\end{tabular}
& \begin{tabular}[c]{@{}c@{}}GNU\\ {\small GCC}\end{tabular}
& \begin{tabular}[c]{@{}c@{}}HPE\\ {\small CCE}\end{tabular}
& \begin{tabular}[c]{@{}c@{}}Intel\\ {\small ICX/IFX}\end{tabular}
& \begin{tabular}[c]{@{}c@{}}LLVM\\ {\small Clang/Flang}\end{tabular}
& \begin{tabular}[c]{@{}c@{}}NVIDIA\\ {\small NVHPC}\end{tabular} \\ \midrule
 & \multicolumn{2}{l}{Runtime Name} & \texttt{\small libomp} & \texttt{\small libomp} & \texttt{\small libomp} & \texttt{\small libomp} & \texttt{\small libgomp} & \texttt{\small libcraymp} & \texttt{\small libomp} & \texttt{\small libomp} & \texttt{\small libnvomp} \\
 & Language Support & C/C++ & \checkmark & \checkmark & \checkmark & \checkmark & \checkmark & \checkmark & \checkmark &\checkmark & \checkmark \\
 &  & Fortran & \checkmark & \checkmark & \checkmark & \checkmark & \checkmark & \checkmark & \checkmark & \checkmark & \checkmark \\
4.0 & Target Offloading & AMD & \checkmark & \checkmark & \checkmark & - & \checkmark & \checkmark & \checkmark & \checkmark & - \\
 &  & Intel & - & - & - & - & - & - & \checkmark & - & - \\
 &  & NVIDIA & \checkmark & \checkmark & \checkmark & - & \checkmark & \checkmark & \checkmark & \checkmark & \checkmark \\
5.0 & \multicolumn{2}{l}{Tool Initialization} & 2.0 & 0.8-0 & 3.5.0 & 20.0 & - & 11.0.0 & 2021.1 & 8.0.0 & 22.7 \\
 & \multicolumn{2}{l}{Target Callback\footnote{\label{note:deprecated-60}This feature is \emph{deprecated} in OpenMP 6.0 \cite{openmpSpec60}, no longer \emph{required} for OMPT compliance.}} & 5.0 & 17.0-3 & 5.7.0 & - & - & 16.0.0 & 2023.2 & 17.0.1 & 22.7 \\
 & \multicolumn{2}{l}{Target Data Op Callback\footref{note:deprecated-60}} & 5.0 & 17.0-3 & 5.7.0 & - & - & 16.0.0 & 2023.2 & 17.0.1 & 22.7 \\
 & \multicolumn{2}{l}{Target Submit Callback\footref{note:deprecated-60}} & 5.0 & 17.0-3 & 5.7.0 & - & - & 16.0.0 & 2023.2 & 17.0.1 & 22.7 \\
 & \multicolumn{2}{l}{Target Map Callback\footref{note:deprecated-60}\footnote{\label{note:optional}Implementation of this feature is \emph{optional} for OMPT compliance.}} & - & - & - & - & - & - & - & - & 22.7 \\
 & \multicolumn{2}{l}{Tracing Interface} & - & 14.0-1 & 5.1.0 & - & - & - & - & - & - \\
5.1 & \multicolumn{2}{l}{Target EMI Callback\footnote{\label{note:ompdataperf-required}This feature is \emph{required} for \texttt{OMPDataPerf}.}} & 5.0 & 17.0-3 & 5.7.0 & - & - & 16.0.0 & 2023.2 & 17.0.1 & 22.7 \\
 & \multicolumn{2}{l}{Target Data Op EMI Callback\footref{note:ompdataperf-required}} & 5.0 & 17.0-3 & 5.7.0 & - & - & 16.0.0 & 2023.2 & 17.0.1 & 22.7 \\
 & \multicolumn{2}{l}{Target Submit EMI Callback} & 5.0 & 17.0-3 & 5.7.0 & - & - & 16.0.0 & 2023.2 & 17.0.1 & 22.7 \\
 & \multicolumn{2}{l}{Target Map EMI Callback\footref{note:optional}} & - & - & - & - & - & - & - & - & 22.7 \\
\bottomrule
\end{tabularx}
\endgroup
\label{tab:ompt_support}
\end{minipage}
\vspace{-1em}
\end{table*}
\subsection{Compilers/Runtimes Supporting OMPT with EMI Callbacks}

    \subsubsection{LLVM Clang/Flang}
    Clang \cite{llvm-clang}, the C/C++ frontend for LLVM \cite{lattner2004llvm, llvm}, provides one of the most complete OpenMP implementations, with Flang's \cite{llvm-f18} support rapidly developing.
    LLVM's \\ \texttt{libomptarget} runtime supports offloading to AMD and NVIDIA GPUs \cite{llvm-19-1-0-OpenMPSupport}. 
    LLVM first implemented OMPT tool initialization and non-target callbacks in 6.0.0 (March 2018) based on the early technical reports (\href{https://github.com/llvm/llvm-project/commit/82e94a593433f36734e2d34898d353a2ecb65b8b}{\texttt{82e94a}}, \href{https://github.com/llvm/llvm-project/commit/cab9cdc2ad8b1c40d99ac07c5a03b8061a7f3868}{\texttt{cab9cd}}, \href{https://github.com/llvm/llvm-project/commit/e1a04b4659a6cc7ef0f777622b38c3da3535b92d}{\texttt{e1a04b}}, \href{https://github.com/llvm/llvm-project/commit/489cdb783a0655ed2d2812a62fda220eb17574dd}{\texttt{489cdb}}) \cite{llvm-6-0-0-OpenMPSupport}.
    LLVM 8.0.0 (March 2019) finalized these features in accordance with the OpenMP 5.0 specification only a few months after its ratification (\href{https://github.com/llvm/llvm-project/commit/0e0d6cdd5862d7c5e28c542fa07d6fadd07f1628}{\texttt{0e0d6c}}, \href{https://github.com/llvm/llvm-project/commit/2b46d30fc7049fc5cc5b9f98df654509bb4d61a2}{\texttt{2b46d3}}).
    Despite Clang's support documentation \cite{llvm-17-0-1-OpenMPSupport, llvm-18-1-0-OpenMPSupport} not indicating compliant 5.0 OMPT callback and 5.1 EMI support until 18.1.0-rc1 (\href{https://github.com/llvm/llvm-project/commit/99ce17b71c41190cd82f4c5382910d71cb673abe}{\texttt{99ce17}}), these features were actually completed in 17.0.0-rc4 with the inclusion of \href{https://github.com/llvm/llvm-project/commit/7e68c9e5c27182b867c1602a1095395089d3039c}{\texttt{7e68c9}} followed by minor bug fixes \href{https://github.com/llvm/llvm-project/commit/4d5feafb9dc0d7e9b12b116f07307085687c2e3d}{\texttt{4d5fea}} and \href{https://github.com/llvm/llvm-project/commit/1d54dc2f75861295aeb99f480aed244dc5cedea5}{\texttt{1d54dc}}.

    \subsubsection{AMD Optimizing C/C++ and Fortran Compilers (AOCC)} \hspace{0pt}\\
    Highly optimized for AMD Zen-based processors, the October 2024 release of AOCC 5.0 \cite{amd-aocc}, is built on LLVM 17.0.1 and is the first version of AOCC to support target and EMI callbacks while prior versions only supported non-target callbacks \cite{amd-aocc-5-0-OpenMPSupport, amd-aocc-4-2-OpenMPSupport}.

    \subsubsection{AMD ROCm}
    ROCm \cite{amd-rocm} is an open-source software stack optimized for AMD GPUs, consisting of compilers, libraries, and tools.
    ROCm contains a downstream LLVM project with additional enhancements for AMD GPUs.
    Target callbacks support was added in 5.7.0 (\href{https://github.com/ROCm/llvm-project/commit/12e5b16d5fe1a003788190a3c102532d79a18e00}{\texttt{12e5b1}}), released September 2023 \cite{amd-rocm-5-7-0-OpenMPSupport}.
    Additionally, in 5.1.0 (\href{https://github.com/ROCm/llvm-project/commit/b999647daa8d1f141261fa99e69d01690437415c}{\texttt{b99964}}), released March 2022, AMD implemented the OMPT tracing interface, a feature not yet available in the latest upstream LLVM project as of writing \cite{amd-rocm-5-1-0-OpenMPSupport}.

    \subsubsection{AMD AOMP}
    AOMP \cite{amd-aomp} is an open-source LLVM-based compiler with improved support for OpenMP on Radeon GPUs.
    AOMP builds on top of, and releases are more frequent than ROCm LLVM.
    Version 14.0-1 (January 2022) implements the OMPT tracing interface \cite{amd-aomp-14-0-1-OpenMPSupport}.
    Version 17.0-3 (July 2023) provides target and EMI callbacks \cite{amd-aomp-17-0-3-OpenMPSupport}.

    \subsubsection{HPE Cray Compiling Environment (CCE)}
    CCE \cite{hpe-cce} provides the Cray Fortran and C/C++ compilers which use the \texttt{libcraymp} and \texttt{libcrayacc} runtimes which can be used as drop-in replacements for LLVM's \texttt{libomp} and \texttt{libomptarget} runtimes or GNU's \texttt{libgomp} runtime.
    The C/C++ compiler is LLVM-based and supports offloading to AMD and NVIDIA GPUs.
    The November 2020 release of CCE 11.0.0 introduced OMPT 5.0 support \cite{hpe-cce-11-0-enhancements}.
    EMI callbacks were implemented in CCE 16.0.0 (May 2023) \cite{sc23-openmp-bof}.

    \subsubsection{Intel oneAPI DPC++/C++ (ICX) and Fortran (IFX) Compilers}
    ICX/IFX are proprietary LLVM-based compilers that have had target and EMI callbacks since version 2023.2 (July 2023) and support offloading to Intel GPUs. Additional plugins can enable offloading to AMD and NVIDIA GPUs.
    Support continues through the latest stable release as of writing, version 2024.2 (June 2024), and is based on LLVM 19 development \cite{intel-icx-OpenMPSupport, intel-dpcpp-changelog}.

    \subsubsection{NVIDIA HPC SDK (NVHPC)}
    NVIDIA’s HPC SDK \cite{nvidia-hpc} provides a subset of OpenMP 5.0 features and can offload to NVIDIA GPUs.
    Support for OMPT callbacks was introduced in version 22.7 (July 2022) as stated in the compiler release notes, although the compiler documentation does not indicate OMPT or EMI callback support even in the latest 24.9 September 2024 release as of writing \cite{nvidia-hpc-22-7-OpenMPSupport, nvidia-hpc-24-9-OpenMPSupport}.
    We tested to confirm which features were functional.
    Unlike LLVM-based compilers, NVHPC requires (re)compiling your code with the { \tt -mp=ompt } flag in order for OMPT-based tools to be invoked.

\subsection{Compilers/Runtimes Supporting OMPT without EMI Callbacks}

    \subsubsection{Arm Compiler for Linux (ACfL)}
    ACfL \cite{arm-acfl} is an LLVM-based compiler with optimizations for 64-bit Armv8-A server-class processors.
    The latest release, 24.04, is based on LLVM 18 but has offloading support disabled \cite{arm-c-OpenMPSupport, arm-fortran-OpenMPSupport}.
    Only non-target OMPT callbacks are supported.
    
\subsection{Compilers/Runtimes without OMPT Support}

    \subsubsection{GNU Compiler Collection (GCC)}
    The latest release of GCC \cite{gnu-gcc} as of writing, version 14.2, released August 2024, supports offloading to AMD and NVIDIA GPUs.
    However, GCC lacks support for OMPT entirely \cite{gnu-libgomp-14-2-OpenMPSupport}.
    This positions GCC behind LLVM-based compilers in terms of OpenMP tooling.

\subsection{Summary of OMPT Adoption Findings}

LLVM-based compilers, such as Clang/Flang, AMD AOCC/\hspace{0pt}ROCm/\hspace{0pt}AOMP, and Intel ICX/IFX, which closely track upstream LLVM developments, have mature, well-tested OMPT support that has seen steady and development to meet the latest OpenMP specifications.
NVIDIA's HPC SDK also achieves a high level of OMPT support and has target related OMPT features, a year prior to LLVM-based compilers.
Although Arm ACfL is based on LLVM, it does not have support for target offloading enabled.
However, it does include non-target-related OMPT support.
GNU GCC is the only compiler infrastructure we found that lacks OMPT support altogether.
Table \ref{tab:ompt_support} compares and summarizes OMPT-related feature support across these compilers and indicates the earliest version we could determine to support each feature.
Additionally, the footnotes of Table \ref{tab:ompt_support} indicate each feature's status in later OpenMP standard revisions, whether each feature is required to be compliant, and whether \texttt{OMPDataPerf} depends on the feature.

\subsection{Recommendations}

Proper implementation documentation of OMPT features by compiler/runtime implementers remains needed.
Among the compilers we examined that claimed OMPT support, \emph{none} clearly and/or correctly specified which exact features (i.e. callbacks/tracing interface) they actually supported.
Additionally, AMD AOMP and ROCm, both based on the same LLVM branch, are the only compilers to support the OMPT tracing interface.
Upstreaming this feature into the mainline LLVM should be considered to promote broader adoption.

For tool designers, the target EMI callbacks, excluding the target map EMI callback, are well supported across runtimes.
If ensuring maximal tool runtime compatibility or support going forward is a priority, tool developers should limit their usage of OMPT target callbacks to the EMI callbacks, again excluding the optional map callback.
Although the non-EMI versions of the same callbacks are currently well supported, they have recently been deprecated in OpenMP 6.0 \cite{openmpSpec60}.

\end{document}